\documentclass[aps,twocolumn,floatfix,superscriptaddress,longbibliography]{revtex4-1}

\usepackage{graphicx}
\usepackage{indentfirst}
\usepackage{latexsym}
\usepackage{multirow}
\usepackage{tabls}
\usepackage{epsfig}
\usepackage{multirow}
\usepackage{subfigure}
\usepackage{comment}
\usepackage{hyperref}
\usepackage{float}
\usepackage{color}
\usepackage[usenames,dvipsnames]{xcolor}
\usepackage{amssymb}
\usepackage{amsfonts}
\usepackage{amsmath}
\usepackage{bm}
\usepackage{mathrsfs}

\newcommand{\Msun}{\ensuremath{M_{ \odot }}}

\mathchardef\mhyphen="2D
\newcommand{\CSt}{{\mbox{\tiny CS2}}}

\begin{document}

\title{Testing the ``no-hair'' property of black holes with\\X-ray observations of accretion disks}

\def\addCambridge{Institute of Astronomy, Madingley Road, Cambridge, CB30HA, United Kingdom}

\author{Christopher J. Moore}
\email{cjm96@ast.cam.ac.uk}
\affiliation{\addCambridge} 

\author{Jonathan R. Gair}
\email{jrg23@ast.cam.ac.uk} 
\affiliation{\addCambridge}

\date\today

\begin{abstract}
Accretion disks around black holes radiate a significant fraction of the rest mass of the accreting material in the form of thermal radiation from within a few gravitational radii of the black hole ($ r \lesssim 20 G M / c^{2}$). In addition, the accreting matter may also be illuminated by hard X-rays from the surrounding plasma which adds fluorescent transition lines to the emission. This radiation is emitted by matter moving along geodesics in the metric, therefore the strong Doppler and gravitational redshifts observed in the emission encode information about the strong gravitational field around the black hole. In this paper the possibility of using the X-ray emission as a strong field test of General Relativity is explored by calculating the spectra for both the transition line and thermal emission from a thin accretion disk in a series of parametrically deformed Kerr metrics. In addition the possibility of constraining a number of known black hole spacetimes in alternative theories of gravity is considered.
\end{abstract}


\maketitle

\section{Introduction}
General Relativity (GR) has been extensively tested in the weak field regime. The famous classical tests of GR began before the theory was even fully formulated with the explanation of the anomalous precession of the perihelion of mercury \cite{1915SPAW.......831E}. Only a few years later further confirmation came with the first measurements of the gravitational deflection of light during the 1919 solar eclipse \citep{1920RSPTA.220..291D}. More recently, but no less famously, the confirmation by \cite{1975ApJ...195L..51H} that the loss of energy from the binary pulsar PSR1913+16 was in the manner predicted by the theory won the authors the 1993 Nobel prize in physics~\footnote{\url{http://www.nobelprize.org/nobel_prizes/physics/laureates/1993/index.html}}. Interspersed among these landmark experiments GR has been the subject of a constant stream of increasingly stringent tests; to date no deviations from the predictions of GR have been detected. For a review of experimental tests of GR see \cite{lrr-2006-3}. 

All of the tests mentioned above concern phenomena in weak gravitational fields; it has proved much more difficult to devise similarly stringent tests of GR in the strong gravitational field. This is due in part to the immense difficulty associated with measuring the motion of bodies on the small gravitational length scale associated with black holes (a few kilometers for a solar mass black hole) at astrophysical distances. If such measurements could be made and it were possible to accurately track the motion of a test particle around an astrophysical black hole it would become possible to place constraints on the strong field spacetime metric. It is a prediction of GR, together with the no-hair theorem, that the spacetime metric around an astrophysical black holes is described by the famous Kerr solution \citep{PhysRevLett.11.237}. Alternative theories of gravity predict the existence of different solutions (examples of known solutions considered in this paper include the Kehagias-Sfetsos solution \citep{2009PhLB..678..123K} to Ho\v{r}ava gravity \citep{2009PhRvL.102p1301H,2009PhRvD..79h4008H} and the slowly rotating solution \citep{2009PhRvD..79h4043Y} to dynamical Chern-Simons gravity \citep{2003PhRvD..68j4012J}). If a deviation from the Kerr solution was detected it could indicate a failure of GR in the strong field regime. Alternatively, it could indicate a failure of the no-hair theorem and raise the possibility of exotic compact objects within GR. It should be noted that there are other ways in which alternative theories of gravity may differ from GR besides changes to the metric around a black hole; for example, scalar-tensor theories of gravity (of which Brans-Dicke theory \citep{PhysRev.124.925} is perhaps the best known example) generically predict the existence of addition gravitational wave polarisation states \citep{lrr-2006-3}. In this paper however we restrict our attention to testing the spacetime metric around a black hole.

Tests of the metric around a black hole may be possible in the near future with gravitational wave observations, such as those from ground-based detectors like advanced LIGO and advanced VIRGO \citep{2010CQGra..27h4006H,Acernese2009}. For example, the possibility of using a network of ground-based gravitational wave detectors observing the coalescences of binary neutron stars to constrain the deviation of a post-Newtonian coefficient from the predicted GR value was considered in \cite{2014PhRvD..89h2001A}. Further into the future, low-frequency space-based gravitational wave observatories such as eLISA \citep{TheGravitationalUniverse} will offer the possibility of probing the gravitational field around supermassive black holes. The possibility of using observations of the inspiral of a stellar mass compact object into a supermassive black hole to constrain the anomolous quadrupole moment of the massive black hole was considered by \cite{1997PhRvD..56.1845R}, and later by \cite{2007PhRvD..75d2003B}. These gravitational wave observations are ideally suited to such tests as they offer the ability to accurately track the orbital phase of the two bodies all the way up to and including the merger. However, as gravitational wave observation are not yet available it is interesting to ask if strong field tests are possible using current electromagnetic observations.  

Accretion disks provide an ideal candidate for such tests, because X-ray irradiation of matter in the inner regions of the disk ($r \lesssim 20M$) imprints characteristic features upon the X-ray spectra, in particular the fluorescent K$\alpha$ Iron line (rest energy 6.38keV). As this emission originates from so close to the Black Hole (BH)  it is strongly distorted by gravitational and Doppler shifting producing the characteristic `two-horned' profiles observed in nature \cite{1995Natur.375..659T}. Furthermore the light, once emitted, is strongly gravitationally lensed as it propagates through the spacetime, altering the observed spectra and imprinting upon it extra information about the strong field metric. Indeed the large width of these lines is evidence for the existence of highly spinning black holes \citep{1996MNRAS.279..837I}; for a recent review of black hole spin measurements using X-ray emission see \cite{2011CQGra..28k4009M}. In addition, viscous torques present in the disk dissipate energy causing the material to gradually spiral inwards, this energy is radiated locally in the form of thermal emission. This thermal radiation is subjected to the same strong-field gravitational effects as the line emission, and so may also be used as a probe of the metric. 

The approach of this paper is to consider Iron line and thermal emission in a large class of parametrically deformed Kerr black holes. The deformed black hole spacetimes are referred to as ``bumpy black holes'' and the individual deformation parameters as ``bumps''. The spacetimes all have the property that if all the bump parameters are simultaneously set to zero then the Kerr solution is recovered. The advantage of this approach, compared with considering the emission in a small number of known alternative black hole solutions, is that one is able to consider a wide range of different deformations simultaneously and identify particular ``bumps'' which are especially easy or difficult to constrain. Furthermore, even if a particular alternative black hole is not contained within this class it may be hoped that the disk spectrum in this new metric would have a significant overlap with the spectrum of a black hole in the class; and therefore a deviation from GR would still be detectable. The family of bumpy black holes used for this purpose was the class of metrics constructed by \cite{2011PhRvD..83j4027V}; these have the property that the perturbed spacetimes possess a fourth constant of motion (analogous to the Carter constant \citep{PhysRev.174.1559} in Kerr spacetime). In addition to this very general familiy of bumpy black holes the results are put into context by comparing them with the bounds it is possible to place on some known black hole solutions; in particular both the linear \cite{2009PhRvD..79h4043Y} and quadratic \citep{2012PhRvD..86d4037Y} in spin solutions to weakly coupled dynamical Chern-Simons gravity, and the slowly rotating Kehagias-Sfetsos black hole \citep{2010EPJC...70..367L} were considered.

In this paper we begin in Sec.\ \ref{sec:spacetimes} by describing the family of bumpy black holes that were considered in this paper, we also describe a couple of specific examples of known black hole solutions in alternative theories of gravity which are also considered here. In Sec.\ \ref{sec:emission} the necessary theory for calculating both the thermal and Iron line emission from an accretion disk in a general stationary, axisymmetric spacetime is described. Sec.\ \ref{sec:analysis} describes the methods used for data analysis is this paper and the method used for estimating the bounds it will be possible to place on the different deformation, or ``bump'', parameters. The results are presented in Sec.\ \ref{sec:results} and finally a discussion and concluding remarks are given in Sec.\ \ref{sec:discussion}. Throughout this paper natural units, where $G=c=k_{B}=1$, are used.

\section{Bumpy black hole spacetimes}\label{sec:spacetimes}
A metric that represents a rotating black hole is required to be stationary and axisymmetric (i.e. there exists a timelike Killing vector $\partial/\partial t$ and a spacelike Killing vector $\partial/\partial\phi$), invariant under simultaneous inversion of the $\phi$ and $t$ coordinates, and reflection symmetric about the equatorial plane. A sufficiently general metric which captures all these properties is given by \citep{Chandrasekhar:579245}
\begin{equation}\label{eq:generalmetric} \textrm{d}s^{2}=g_{tt}\,\textrm{d}t^{2} + g_{t\phi}\,\textrm{d}t\,\textrm{d}\phi + g_{rr}\,\textrm{d}r^{2} + g_{\theta\theta}\,\textrm{d}\theta^{2} + g_{\phi\phi}\,\textrm{d}\phi^{2}\, , \end{equation}
where the metric coefficients depend only on the radial and polar coordinates $r$ and $\theta$. In fact Eq.\ \ref{eq:generalmetric} retains a considerable degree of gauge freedom, however, it is sufficient for our present purpose. 

As our interest here is in testing the hypothesis that the metric is the Kerr solution, and it is known that the Kerr solution is an excellent description in the weak field, it is natural to expand the metric as
\begin{equation} g_{\mu \nu}=g^{\textrm{Kerr}}_{\mu \nu}+\epsilon h_{\mu \nu} \, ,\end{equation}
where $\epsilon \ll 1$ and $h_{\mu\nu}\rightarrow 0$ as $r\rightarrow\infty$. The background Kerr solution is given by,
\begin{eqnarray} &g^{\textrm{Kerr}}_{tt}=-\left( 1-\frac{2Mr}{\rho^{2}} \right)\, , \;  g^{\textrm{Kerr}}_{t\phi}=\frac{-2M^{2}ar}{\rho^{2}}\sin ^{2} \theta \, ,\nonumber \\
&g^{\textrm{Kerr}}_{rr}=\frac{\rho^{2}}{\Delta}\, , \; g^{\textrm{Kerr}}_{\theta \theta}=\rho^{2} \, , \; g^{\textrm{Kerr}}_{\phi \phi}=\frac{\Sigma^{2}}{\rho^{2}}\sin ^{2} \theta \, , \end{eqnarray}
for a BH with a mass $M$ and dimensionless spin $a$, in which
\begin{eqnarray} &\rho^{2}=r^{2}+a^{2}M^{2}\cos ^{2} \theta \, , \; \Delta = r^{2}\left( 1-\frac{2M}{r} \right) +M^{2}a^{2} \, ,\\
&\textrm{and}\quad \Sigma^{2} = \left( r^{2}+M^{2}a^{2} \right)^{2} - M^{2}a^{2}\Delta \sin ^{2} \theta \, \nonumber .\end{eqnarray}
From here on dimensionless units with $M=1$ are used.

Geodesic motion with four-momentum $p^{\mu}$ in a metric with the symmetries outlined thus far would posses three constants of motion; energy ${E=-p^{\mu}(\partial/\partial t)_{\mu}}$, z-component of angular momentum ${L=p^{\mu}(\partial/\partial \phi)_{\mu}}$, and particle rest mass ${m^{2}=-p^{\mu}p_{\mu}}$. However the Kerr solution also possess an additional constant of motion related to the existence of a second rank Killing tensor, ${C=p^{\mu}p^{\nu}\xi_{\mu\nu}}$. In place of $C$ the Carter constant is often defined as ${Q=C-(L-aE)^{2}}$ \citep{PhysRev.174.1559}. The existence of the addition constant of motion ensures that geodesic motion in the spacetime is separable and tri-periodic (i.e.\ with a well defined frequency associated with motion in each of the $r$, $\theta$ and $\phi$ coordinates). Although it is not clear that it should be required of any alternative to the Kerr solution to possess these properties, they are sufficiently appealing that it is worth considering the possibility carefully.

The most general form of metric perturbation, $h_{\mu\nu}$, that could be added to the Kerr solution such that the resultant metric also admitted a second rank Killing tensor and hence a Carter-like constant (at least to ${\cal{O}} (\epsilon ^{2})$ in the perturbation) was considered by \cite{2011PhRvD..83j4027V}. The metric perturbation was also required to tend to zero in the limit $r\rightarrow\infty$ faster than $r^{-2}$ so that all weak field tests are still satisfied. A similar solution was also found earlier by \cite{1979GReGr..10...79B}. Working in Boyer-Linquist-like coordinates \cite{2011PhRvD..83j4027V} found a series of differential relations that must be satisfied by the metric perturbation components. The only non zero components of the perturbation are $\left\{ h_{tt}, h_{t \phi }, h_{rr}, h_{ \phi \phi } \right\}$; in particular $h_{\theta\theta}$ vanishes. In subsequent work \cite{2011PhRvD..84f4016G} expansions for the metric perturbation components in powers of $1/r$ were found,
\begin{equation} h_{\mu \nu}=\sum_{n}h_{\mu \nu, n}\left( \frac{1}{r} \right)^{n} \end{equation}
where the coefficients $h_{\mu \nu , n}$ are functions of $\theta$ only. The leading order non-zero coeffeicients are given in Eq.\ \ref{eq:B2metriccomps} and all the coefficients up to ${\cal{O}}(1/r^{5})$ are reproduced in Appendix \ref{subsec:BNcomponents};
\begin{eqnarray}\label{eq:B2metriccomps}
&h_{tt,2}&=\gamma_{1,2}+2\gamma_{4,2}-2a\gamma_{3,1}\sin^{2}\theta \, , \nonumber \\
&h_{rr,2}&=-\gamma_{1,2}\, ,\nonumber\\
&h_{t\phi,2}&=-M\sin^{2}\theta\left[ \gamma_{3,3}+a\left(\gamma_{1,2}+\gamma_{4,2} \right) +a^{2}\gamma_{3,1} \right] \, , \nonumber \\
&h_{\phi\phi,0}&=2M^{2}a\gamma_{3,1}\sin^{4}\theta\, .
\end{eqnarray}
Note that the leading order correction enters at a lower order in the $h_{\phi\phi}$ component. The components $h_{tt}$ and $h_{rr}$ are dimensionless, whilst $h_{t\phi}$ has units of length and $h_{\phi\phi}$ has units of length squared; reflecting the dimensions of the components of $g_{\mu\nu}$.

At low order in the expansion the metric perturbation is fully characterised by a small number of coefficients, $\gamma_{i,j}$. Adopting the notation of \cite{2011PhRvD..84f4016G} the metric perturbation up to ${\cal{O}}(1/r^{2})$ is given by the four constants ${\cal{B}}_{2}=\left(\gamma_{1,2},\gamma_{3,1},\gamma_{3,3},\gamma_{4,2}\right)$ (see Eq.\ \ref{eq:B2metriccomps}); up to ${\cal{O}}(1/r^{3})$ it is given by the 7 constants ${\cal{B}}_{2}\cup{\cal{B}}_{3}$, where ${\cal{B}}_{3}=(\gamma_{1,3}, \gamma_{3,4}, \gamma_{4,3})$; up to ${\cal{O}}(1/r^{4})$ it is given by the 10 constants ${\cal{B}}_{2}\cup{\cal{B}}_{3}\cup{\cal{B}}_{4}$, where ${\cal{B}}_{4}=(\gamma_{1,4}, \gamma_{3,5}, \gamma_{4,4})$; up to
${\cal{O}}(1/r^{5})$ it is given by the 13 constants ${\cal{B}}_{2}\cup{\cal{B}}_{3}\cup{\cal{B}}_{4}\cup{\cal{B}}_{5}$, where ${\cal{B}}_{5}=(\gamma_{1,5}, \gamma_{4,5}, \gamma_{3,6})$.

Throughout this paper various ${\cal{B}}_{N}$ limits will be referred to, these correspond to setting all the parameters $\gamma_{i,j}$ to zero except for those in the set ${\cal{B}}_{N}$, which are treated as independent free parameters. This greatly reduces the number of free parameters and allows for a systematic way of examining perturbations to the Kerr metric. Because the Kerr metric is known to be an excellent approximation at large radii it is natural to expect any deviation from the Kerr solution to show up, initially at least, at lowest order in $1/r$. This is the justification for examining each of the ${\cal{B}}_{N}$ limits separately. In general, it would be possible for any combination of the $\gamma_{i,j}$ to be be none-zero. In this paper results are presented for the ${\cal{B}}_{2}$, ${\cal{B}}_{3}$, ${\cal{B}}_{4}$ and ${\cal{B}}_{5}$ metrics.

\subsection{Known black hole solutions}
In addition to the very general family of perturbed metrics described above, it is also useful to consider some examples of known black hole solutions in alternative theories of gravity. In this section a few such solutions are listed. One of these spacetimes (the linear in spin solution to dynamical Chern-Simons gravity, Sec.\ \ref{subsec:CS1}) is a member of the family of solutions described in Sec.~\ref{sec:spacetimes}; it can be obtained by a particular choice of the constants $\gamma_{i,j}$. 

\subsubsection{The Kehagias Sfetsos metric}\label{subsec:KS}
A spherically symmetric black hole solution to Ho\v{r}ava gravity \citep{2009PhRvL.102p1301H,2009PhRvD..79h4008H} was found by \cite{2009PhLB..678..123K}. This was generalised to a slowly rotating solution by \cite{2010EPJC...70..367L}. Accretion disk signatures for this type of black hole have been considered previously by \cite{2011CQGra..28p5001H}. The metric is
\begin{eqnarray}\label{eq:KSmetric} 
&\textrm{d}s^{2}_{\textrm{KS}}&=-f_{\textrm{KS}}(r)\textrm{d}t^{2}+\frac{\textrm{d}r^{2}}{f_{\textrm{KS}}(r)}+r^{2}\textrm{d}\Omega^{2}-\frac{4a\sin^{2}\theta}{r}\textrm{d}\phi\textrm{d}t\; , \nonumber\\
&&\textrm{where }\, f_{\textrm{KS}}(r)=1+\omega r^{2}\left(1-\sqrt{1+\frac{4}{\omega r^{3}}}\right)\,, \nonumber \\
&&\textrm{and }\,\textrm{d}\Omega^{2}=\textrm{d}\theta^{2}+\sin^{2}\theta\textrm{d}\phi^{2}\, .\end{eqnarray}
In the limit $\omega \rightarrow \infty$ the slowly rotating limit of the Kerr metric is recovered. In order to avoid a naked singularity at the origin an extra constraint is needed, $\omega M^{2} \geq \frac{1}{2}$. For the remainder of this paper $\omega$ is made dimensionless by multiplying by $M^{2}$, and we choose to work with the small parameter $Y\equiv 1/(\omega M^{2})\ll 1$. For the remainder of this paper the metric in Eq.\ \ref{eq:KSmetric} will be refered to as the KS metric.

\subsubsection{The Chern Simons metric linear in spin}\label{subsec:CS1}
Dynamical Chern Simons (CS) modified gravity \cite{2003PhRvD..68j4012J} is a parity violating theory of gravity constructed by adding a Pontryagin invariant term to the action. As the Schwarzschild solution is spherically symmetric and has even parity it remains a solution to the modified CS field equations; however, the Kerr solution does not have even parity and fails to satisfy these equations. No complete rotating black hole solution is known in the CS theory, however perturbative solutions in the spin and CS coupling constant have been found analytically. The rapidly rotating case was considered numerically by \cite{2014arXiv1407.2350S}, here we focus on the slowly rotating solutions.

The slowly rotating black hole solution (linear in spin) to dynamical Chern-Simons gravity was found by \cite{2009PhRvD..79h4043Y};
\begin{equation}\label{eq:CS1metric} \textrm{d}s^{2}_{\textrm{CS}1}=\textrm{d}s^{2}_{\textrm{Kerr}}+\frac{5}{4}\zeta\chi\frac{1}{r^{4}}\left(1+\frac{12}{7r^{2}}+\frac{27}{10r^{2}}\right)\textrm{d}t\textrm{d}\phi \, .\end{equation}
Accretion disk signatures for this type of black hole have been considered previously by \cite{2010CQGra..27j5010H}. For the remainder of this paper the metric in Eq.\ \ref{eq:CS1metric} will be refered to as the CS1 metric.

\subsubsection{The Chern Simons metric quadratic in spin}\label{subsec:CS2}
A slowly rotating black hole solution (quadratic in spin) to dynamical Chern-Simons gravity was found by \cite{2012PhRvD..86d4037Y}. As far as the authors are aware disk emission in this metric has not been considered before. The metric is
\begin{equation}\label{eq:CS2metric} \textrm{d}s^{2}_\CSt = \textrm{d}s^{2}_{\textrm{Kerr}}+\delta\left(g_{\mu\nu}^\CSt \right)\textrm{d}x^{\mu}\textrm{d}x^{\nu}\, \end{equation}
and expressions for the components $\delta\left(g_{\mu\nu}^\CSt \right)$ are given in Appendix \ref{subsec:CS2components}. For the remainder of this paper the metric in Eq.\ \ref{eq:CS2metric} will be referred to as the CS2 metric.

Unlike the KS metric, both the CS1 and CS2 metrics recover the Schwarzschild solution in the limit $a\rightarrow 0$. This is associated with the fact that dynamical Chern Simons modified gravity is a parity violating theory of gravity and hence the spherically symmetric Schwarzschild solution in GR, which is parity even, remains a solution of the modified theory.

\section{Disk emission in bumpy black holes spacetimes}\label{sec:emission}
The theory of thin accretion disks in the steady state was developed by \cite{1973A&A....24..337S} in the Newtonian case and generalised by \cite{1974ApJ...191..499P} to the general relativistic case. It is assumed that the disk lies in the equatorial plane with material moving in (approximately) circular geodesic orbits, with a small radial accretion velocity superposed on the circular motion. In addition, it is assumed that the disk is thin (i.e. at a radius $r$ the thickness $2h$ satisfies $h\ll r$; the disk height may depend on radius, $h=h(r)$, but for notational simplicity this dependence is suppressed here) and is in a steady state with a constant accretion rate $\dot{M}_{0}$. The disk extends from the innermost stable circular orbit (ISCO) of the spacetime to some finite outer radius (strictly, to be consistent with the steady state assumption, the disk must extend indefinitely; here the more conservative approach of leaving the outer radius as a free parameter is taken). When the accreting material reaches the ISCO it plunges quickly into the BH without radiating any further or interacting viscously with the material at larger radii. The fact that the material plunges quickly after crossing the ISCO provides the physical motivation for imposing a zero torque boundary condition at the inner edge of the disk. This allows the differential equation governing the radial dependence of the flux to be integrated; see Eq.\ \ref{eq:radialfluxmain}. 

The model described above is often referred to as the standard relativistic model of accretion disks, and has been extensively studied for the Kerr geometry. Here it is necessary to retain sufficient generality to perform the calculations for any of the bumpy black holes described in Sec.\ \ref{sec:spacetimes}. Since the disk is thin and assumed to lie in the equatorial plane we may switch from using Boyer-Lidquist polar-like coordinates to cylindrical-like coordinates. The metric near the equatorial plane is given by
\begin{equation}\label{eq:metriccylindrical} \textrm{d}s^{2} = g_{tt}\textrm{d}t^{2}+2g_{t\phi}\textrm{d}t\textrm{d}\phi +g_{rr}\textrm{d}r^{2} + g_{\phi\phi}\textrm{d}\phi ^{2} + g_{zz} \textrm{d}z^{2}\, , \end{equation}
where the metric components depend only on $r$; corrections that depend on $z$ enter at second order and are neglected because the disc is assumed to be thin. By a rescaling of the $z$ coordinate the metric component $g_{zz}$ may be set to unity without any loss of generality.

The metric in Eqs.\ \ref{eq:generalmetric} and \ref{eq:metriccylindrical} does not depend on the timelike or azimuthal coordinates, so the corresponding covariant components of the four-momentum are conserved, $p_{t}=-E$ and $p_{\phi}=L$. Using the metric to raise the indices on these momentum components gives the first two geodesic equations in first order form, where a tilde denotes an orbital quantity per unit mass of the test particle,
\begin{eqnarray}\label{eq:timegeo} 
\frac{\textrm{d}t}{\textrm{d}\tau} 	=\frac{\tilde{E}g_{\phi \phi}+\tilde{L}g_{t \phi}}{g_{t \phi}^{2}-g_{tt}g_{\phi \phi}}\, , \nonumber \\ 
\frac{\textrm{d}\phi}{\textrm{d}\tau} 	=-\frac{\tilde{E}g_{t\phi}+\tilde{L}g_{tt}}{g_{t \phi}^{2}-g_{tt}g_{\phi \phi}}\, ,
\end{eqnarray}
where $\tau$ is the proper time along the worldline of the orbiting test particle.

Because the material in the disk is orbiting in the equatorial plane the vertical component of its four-velocity vanishes, $\textrm{d}z/\textrm{d}\tau=0$. Therefore, the third and final geodesic equation may be conveniently obtained from the normalisation condition of the four velocity;
\begin{equation}
\label{eq:norm4velocity}g_{\mu\nu}\frac{\textrm{d}x^{\mu}}{\textrm{d}\tau}\frac{\textrm{d}x^{\nu}}{\textrm{d}\tau}\equiv g_{\mu\nu}u^{\mu}u^{\nu}=-1 \, , 
\end{equation}
\begin{equation}
\label{eq:radialgeo}\Rightarrow \left(\frac{\textrm{d}r}{\textrm{d}\tau}\right)^{2} = \frac{V_{\textrm{eff}}(r)}{g_{rr}} \, , 
\end{equation}
where,
\begin{equation} V_{\textrm{eff}}(r)= \frac{\tilde{E}^{2}g_{\phi \phi}+2\tilde{L}\tilde{E}g_{t\phi}+\tilde{L}^{2}g_{tt}}{g_{t\phi}^{2}-g_{tt}g_{\phi\phi}}-1 \, . \end{equation} 
From Eq.\ \ref{eq:radialgeo} it can be seen that the conditions for stable circular orbits are $V_{\textrm{eff}}(r)\nobreak=\nobreak V_{\textrm{eff},r}(r)\nobreak=\nobreak0$, where a comma in the subscript denotes a partial derivative with respect to all subsequent indices. These two conditions yield expressions for the specific energy and angular momentum per unit mass of a particle on a circular, equatorial geodesic;
\begin{eqnarray} 
\tilde{E} = -\frac{g_{tt}+g_{t\phi}\Omega}{\sqrt{-\left( g_{tt}+2g_{t\phi}\Omega+g_{\phi\phi}\Omega^{2} \right)}} \label{eq:En} \\
\tilde{L} = \frac{g_{t\phi}+g_{\phi\phi}\Omega}{\sqrt{-\left( g_{tt}+2g_{t\phi}\Omega+g_{\phi\phi}\Omega^{2} \right)}}\; . \label{eq:Lz}
\end{eqnarray}
Combining Eqs.\ \ref{eq:timegeo} with Eqs.\ \ref{eq:En} and \ref{eq:Lz} gives an expression for the coordinate angular velocity of the particle,
\begin{equation}\label{eq:omega} \Omega = \frac{\textrm{d}\phi}{\textrm{d}t} = \frac{-g_{t\phi ,r}+\sqrt{ \left( g_{t\phi,r} \right)^{2} - g_{tt,r} g_{\phi\phi ,r} }}{g_{\phi \phi , r}} \; . \end{equation}

In the metrics of interest in this paper (i.e.\ the Kerr metric and small perturbations from it) there exists an ISCO. The radius of the ISCO may be found by solving $V_{\textrm{eff}, r r}(r)=0$ simultaneously with $V_{\textrm{eff}}(r)\nobreak=\nobreak V_{\textrm{eff},r}(r)\nobreak=\nobreak0$. This equation admits the following analytic solution in the Kerr case \citep{1972ApJ...178..347B};
\begin{eqnarray}\label{eq:KerrISCO}
&Z_{1}=1+\left(1-a^{2}\right)^{1/3}\left((1+a)^{1/3}+(1-a)^{1/3}\right) \; ,\nonumber\\
&Z_{2}=\left(3a^{2}+Z_{1}^{2}\right)^{1/2} \; , \nonumber\\
&r_{isco}= 3+Z_{2}-\left((3-Z_{1})(3+Z_{1}+2Z_{2})\right)^{1/2}\, .
\end{eqnarray}
For the general bumpy black holes discussed in Sec.\ \ref{sec:spacetimes} the radius for the ISCO must be found numerically. However, if the spacetime is characterised by a small deformation from Kerr, then the radius of the ISCO may be expanded perturbatively in the bump parameter, $\epsilon$,
\begin{equation}\label{eq:iscoshift} r_{isco}=r^{\textrm{Kerr}}-\epsilon \frac{ \left.\frac{\textrm{d}V_{\textrm{eff},rr}(r,\epsilon)}{\textrm{d}\epsilon}\right|_{r=r^{\textrm{Kerr}}_{isco},\epsilon=0} }{\left.\frac{\textrm{d}V_{\textrm{eff},rr}(r,\epsilon)}{\textrm{d}r}\right|_{r=r^{\textrm{Kerr}}_{isco},\epsilon=0}} +{\cal{O}}\left(\epsilon^{2}\right)\; .\end{equation}
If the metric is characterised by more than one bump parameter, $\epsilon_{i}$, as is the case with the ${\cal{B}}_{N}$ spacetimes described in Sec.\ \ref{sec:spacetimes} then the following replacement should be made in Eq.\ \ref{eq:iscoshift},
\begin{eqnarray} &\epsilon\left.\frac{\textrm{d}V_{\textrm{eff},rr}(r,\epsilon)}{\textrm{d}\epsilon}\right|_{r=r_{isco}^{\textrm{Kerr}},\epsilon=0} \rightarrow\sum_{i}\epsilon_{i}\left.\frac{\textrm{d}V_{\textrm{eff},rr}(r,\epsilon_{i})}{\textrm{d}\epsilon_{i}}\right|_{r=r^{\textrm{Kerr}}_{isco},\epsilon_{i}=0} .\nonumber\\
&\end{eqnarray}

From Eq.\ \ref{eq:omega}, and the fact that the orbit is circular and equatorial, the four-velocity of a small fluid element of the disk orbiting the black hole is given by
\begin{equation}\label{eq:circorbfourvel} \left( \frac{\textrm{d}t}{\textrm{d}\tau} , \frac{\textrm{d}r}{\textrm{d}\tau} , \frac{\textrm{d}\phi}{\textrm{d}\tau} , \frac{\textrm{d}z}{\textrm{d}\tau} \right) = \left(u^{t} , u^{r} , u^{\phi} , u^{z} \right) = u^{t} \left( 1,0,\Omega , 0 \right) \, ,\end{equation}
where $u^{t}$ may be found from the normalisation condition on the four velocity in Eq.\ \ref{eq:norm4velocity},
\begin{equation}\label{eq:four:vel} u^{t} = \frac{1}{\sqrt{-\left( g_{tt}+2\Omega g_{t\phi} +\Omega^{2}g_{\phi\phi} \right)}} \; .\end{equation}

Viscous forces in the disk cause the orbiting material to dissipate energy and gradually move inwards to smaller radii. An expression for the radial dependence of the flux of energy from the disk was derived in \cite{1974ApJ...191..499P}; for completeness this derivation is reproduced in Appendix \ref{app:b}. The resulting expression for the radial dependence of the flux is,
\begin{equation}\label{eq:radialfluxmain} F(r)=\frac{-\dot{M}_{0}\Omega_{,r}}{4\pi \sqrt{-{\bf{g}}}\left( \tilde{E}-\Omega \tilde{L} \right)^{2}} \int_{r_{\textrm{isco}}}^{r}\left(\tilde{E}-\Omega \tilde{L}\right) \tilde{L}_{,r}\, \textrm{d}r \; ,\end{equation}
where $\dot{M}_{0}$ is the accretion rate and ${\bf{g}}$ is the metric determinant. Since the disk is in thermodynamic equilibrium, and not heating up or cooling down, this flux of energy is radiated away from the disk in the form of a thermal distribution of photons.

\subsection{Line emission}\label{subsec:line}
\begin{figure*}[t]
 \centering
 \includegraphics[trim=0cm 0cm 0cm 0cm, width=0.9\textwidth]{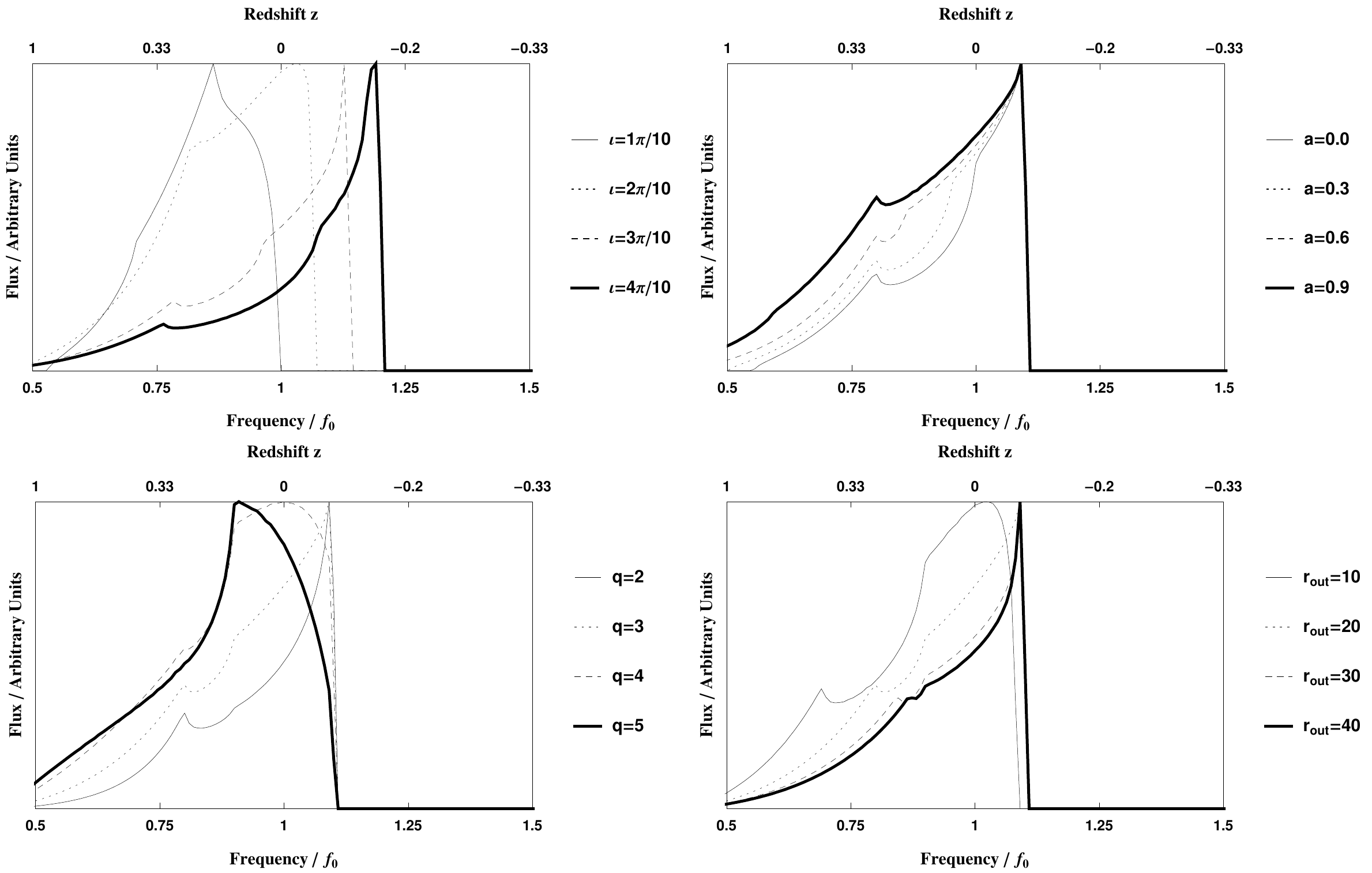}
 \caption{The dependence of the Kerr Iron line profile on various parameters. All spectra were normalised to the same peak flux. Unless otherwise stated in the figure all parameters were set to their fiducial values; $a=0.5$, $\iota=\pi/4$, $q=3$ and $r_{\textrm{out}}=30$.}
 \label{fig:KerrLine}
\end{figure*}

In addition to the flux of thermal radiation a higher energy power-law component of hard X-ray photons is also observed \citep{1994MNRAS.269L..55Z}. For supermassive black holes with cooler disks this component can in fact dominate over the thermal component. The hard X-ray power-law component is generally believed to be caused by inverse compton scattering of the thermal photons radiated from the disk by the hot surrounding corona. A certain fraction of the hard X-ray photons are radiated back towards the disk surface where, upon incidence, they produce fluorescent transition lines which may also be observed as a third component of the spectrum. One frequently observed line is that due to the Iron (Fe) K$\alpha$ transition, which in its rest frame has an energy of $6.38\,\textrm{keV}$. The combined effects of gravitational redshifting and Doppler boosting broaden this line into the characteristic shapes shown in Fig.\ \ref{fig:KerrLine}.

The observed line profile can be computed as follows. Let $\textrm{d}F_{0}(E_{0})$ be the infinitesimal element of flux at energy $E_{0}$ observed at infinity due to an infinitesimal element of the disk. If $I_{0}(E_{0})$ is the specific intensity (i.e.\ the energy per unit time, per unit area, per unit spectral energy (or frequency), per unit solid angle) in the observer's rest frame then
\begin{equation} \textrm{d}F_{0}(E_{0})=I_{0}(E_{0})d\Xi \; ,\end{equation}
where the element of the disk subtends solid angle element $d\Xi$. If the disk has a specific intensity $I_{e}(E_{e})$ in its own rest frame, given the invariance of $I/E ^{3}$ along the photon's worldline \citep{MTW}, it follows that
\begin{equation}\label{eq:fluxtemp1} \textrm{d}F_{0}(E_{0})=g^{3}I_{e}(E_{e})d\Xi \; , \end{equation}
where $g\equiv1/(1+z)\equiv E_{0}/E_{e}$ is the red-shift factor. For line emission the specific intensity may be approximated as a delta-function, $I_{e}(E_{e})=\epsilon (r_{e},\mu_{e})\delta (E_{e}-E_{int})$, where $E_{int}$ is the energy of the transition line in its rest frame, $\epsilon$ is the emissivity of the material, $r_{e}$ is the radius of the emitting fluid element and $\mu_{e}$ is the cosine of the angle of the emission with respect to the disk normal. Eq.\ \ref{eq:fluxtemp1} together with the delta-function expression for the line specific intensity gives
\begin{equation}\label{eq:fluxtemp2} \textrm{d}F_{0}(E_{0})=g^{4}\epsilon(r_{e},\mu_{e})\delta(E_{0}-gE_{\textrm{int}})d\Xi \; ; \end{equation}
the extra factor of $g$ in Eq.\ \ref{eq:fluxtemp1} compared to Eq.\ \ref{eq:fluxtemp2} comes from the change of argument in the delta-function. Now let $r_{e}$ and $\phi_{e}$ be the plane polar coordinates in the disk of the emitting point (where $\phi_{e}=0$ is along the line of nodes where the disk intersects the observers plane of the sky). In addition let $\alpha$ and $\beta$ be the cartesian coordinates in the observer's plane of the sky (where the $x$ axis appears to lie along the lines $\phi_{e}=0$ and $\phi_{e}=\pi$), and $r_{0}$ be the distance to the source. It follows that the solid angle element is given by $\textrm{d}\Xi = \textrm{d}\alpha \textrm{d}\beta / r_{0}^{2}$, and therefore integrating Eq.\ \ref{eq:fluxtemp2} gives
\begin{equation}\label{eq:fluxtemp3} F_{0}(E_{0})=\frac{1}{r_{0}^{2}} \iint\,\textrm{d}\,\alpha\, \textrm{d}\beta\; g^{4}\epsilon(r_{e},\mu_{e})\delta(E_{0}-gE_{\textrm{int}}) \; . \end{equation}
If light-bending is neglected then the polar coordinates of the emitting point in the disk ($r_{e}$ and $\phi_{e}$) may be related to the cartesian coordinates of the point in the observer's plane of the sky ($\alpha_{e}$ and $\beta_{e}$) via the transformation,
\begin{eqnarray}\label{eq:negligiblelightbenging} 
& \alpha_{e}=r_{e}\cos\phi_{e} \; , \;\; \beta_{e}=r_{e}\sin\phi_{e}\cos \iota\; , \nonumber \\
&\textrm{with Jacobian }\, \frac{\partial (\alpha,\beta)}{\partial (r_{e},\phi_{e})}=r_{e}\cos \iota \, ,\end{eqnarray}
where $\iota$ is the inclination from which the disk is viewed (i.e.\ the angle between the observer's line of sight and the spin axis of the BH). For further discussion of this assumption see Sec.\ \ref{subsec:lightbendingcalcs}. With this in hand we may change the integral in Eq.\ \ref{eq:fluxtemp3} from being over the observer's plane of the sky to being over the disk itself (Dropping the subscript $e$),
\begin{equation}\label{eq:fluxtemp4} F_{0}(E_{0})=\frac{\cos \iota}{r_{0}^{2}}\int_{0}^{2\pi}\int_{r_{isco}}^{r_{out}}r\textrm{d}r \textrm{d}\phi\, g^{4}\epsilon (r,\mu)\delta\left(E_{0}-gE_{int}\right)\, .\end{equation}

Hereafter the emissivity is taken to be a function of $r$ only (no $\mu_{e}$ dependence), and is parameterised as a power law, $\epsilon=r^{-q}$ where $q$ will be referred to as the emissivity index. We have now reduced the problem to finding the red-shift factor as a function of the position in the disk. The red-shift factor is defined by
\begin{equation} g \equiv \frac{E_{0}}{E_{e}} = \frac{p_{\mu}v^{\mu}\big|_{\textrm{observer}}}{p_{\mu}u^{\mu}\big|_{\textrm{emitter}}} \; ,\end{equation}
where $p_{\mu}$ is the four-momentum of the photon linking the emitter to the observer, $v^{\mu}$ is the four-velocity of the observer, and $u^{\mu}$ is the orbital four-velocity of the emitter (see Eq.\ \ref{eq:circorbfourvel}). Since the metric is independent of both the $t$ and $\phi$ coordinates, the corresponding components of the photon's four-momentum are conserved;
\begin{equation} \left[ p_{\mu} \right] = (p_{t},p_{r},p_{\theta},p_{\phi}) = (-E,p_{r},p_{\theta},\Lambda) \, .\end{equation}
If the observer is at infinity and is at rest with respect to the black hole the numerator of Eq.\ \ref{eq:redshiftfac} is simply given by ${p_{\mu}v^{\mu}\small|_{\textrm{observer}}=-E}$. Using the orbital four-velocity from Eq.\ \ref{eq:circorbfourvel} the denominator of Eq.\ \ref{eq:redshiftfac} is ${p_{\mu}u^{\mu}\small|_{\textrm{emitter}}=-Eu^{t}+\Omega u^{t} \Lambda}$, and hence the red-shift factor is given by
\begin{equation}\label{eq:redshiftfac} g=  \frac{1}{u^{t}(1-\Omega \lambda)} \; ,\end{equation}
where $\lambda = \Lambda /E$. Since $\lambda$ is conserved along the photon's worldline it may be evaluated at infinity where the indices on the photon four-momentum may be raised and lowered using the usual Minkowski metric. At large distance the ratio of the contravariant components of the photon's four-momentum gives the azimuthal impact parameter $\alpha$, defined in Eq.\ \ref{eq:negligiblelightbenging},
\begin{equation} \alpha = -\frac{r p^{\phi}}{p^{t}} \Big|_{r\rightarrow \infty} =-\frac{\lambda}{\sin\iota}\; .\end{equation}

\begin{figure}[h]
 \centering
 \includegraphics[trim=0cm 0cm 0cm 0cm, width=0.45\textwidth]{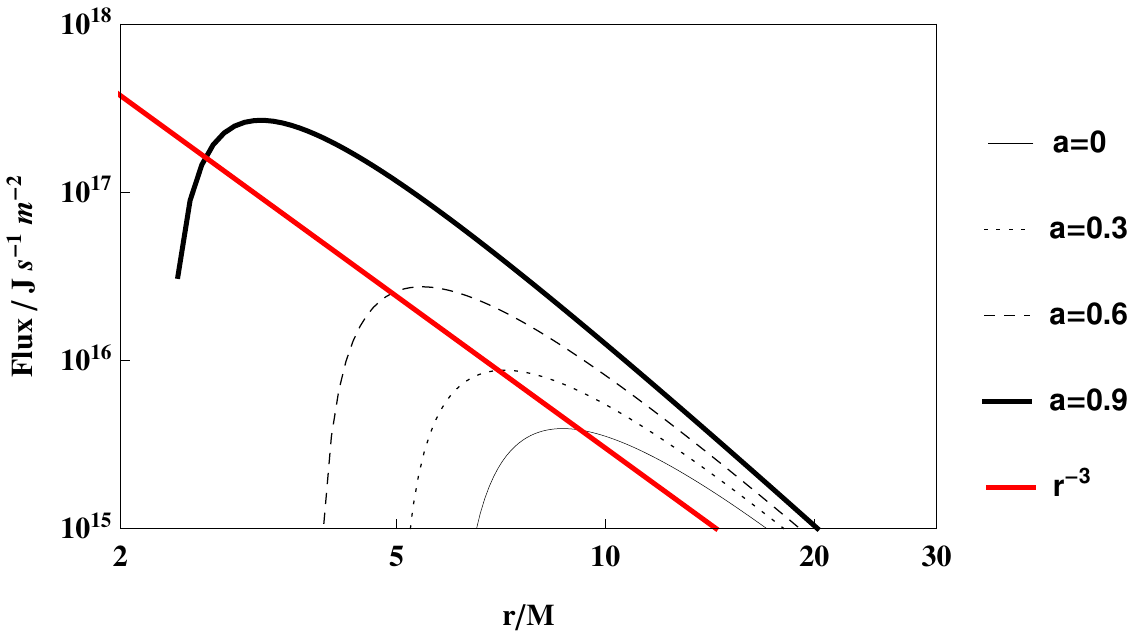}
 \caption{The Flux as a function of radius, for various values of the spin parameter. Also shown is the power law, $(r/M)^{-3}$. All other parameters were set to their fiducial values, $\iota=\pi/4$, $M=\Msun$, $\dot{M}_{0}=10^{-12}\Msun/\textrm{yr}$ and $r_{\textrm{out}}=30$.}
 \label{fig:KerrFlux}
\end{figure}

If light-bending is neglected then the photon linking the emitting point in the disk to the observer at infinity travels on a straight line with impact parameter $\alpha = r\cos\phi $ (see Eq.\ \ref{eq:negligiblelightbenging}). Substituting these results into Eq.\ \ref{eq:redshiftfac} gives the red-shift;
\begin{equation} g\left(r,\phi \right)=\frac{\sqrt{-\left( g_{tt}+2\Omega g_{t\phi}+\Omega^{2}g_{\phi\phi} \right)}}{1+\Omega r \cos \phi \sin \iota} \, . \end{equation}
The flux integral in Eq.\ \ref{eq:fluxtemp4} may now be simply evaluated numerically. This was done initially for the unperturbed Kerr metric in order to reproduce known results (see, for example, \cite{1991MNRAS.250..629K,1991ApJ...376...90L}) and examine the dependence of the spectra on various parameters. The results of these calculations are shown in Fig.\ \ref{fig:KerrLine}. For all calculations in this paper, unless otherwise stated, the disk parameters were set to the following fiducial values; $a=0.5$, $\iota=\pi/4$, $q=3$ and $r_{\textrm{out}}=30$.

It can be seen from Fig.\ \ref{fig:KerrLine} that the profile is strongly sensitive to disk inclination. Qualitatively this is because the orbiting material is moving at relativistic speeds and for large inclinations material on one side of the disk is moving towards the observer while on the other side it is moving away. This produces both a blue and a red-shifted horn to the profile, the blue-shifted horn is always more intense due to the relativistic beaming effect which enhances the intensity of light emitted in the direction of travel (the headlight effect). 

\begin{figure*}[t]
 \centering
 \includegraphics[trim=0cm 0cm 0cm 0cm, width=0.9\textwidth]{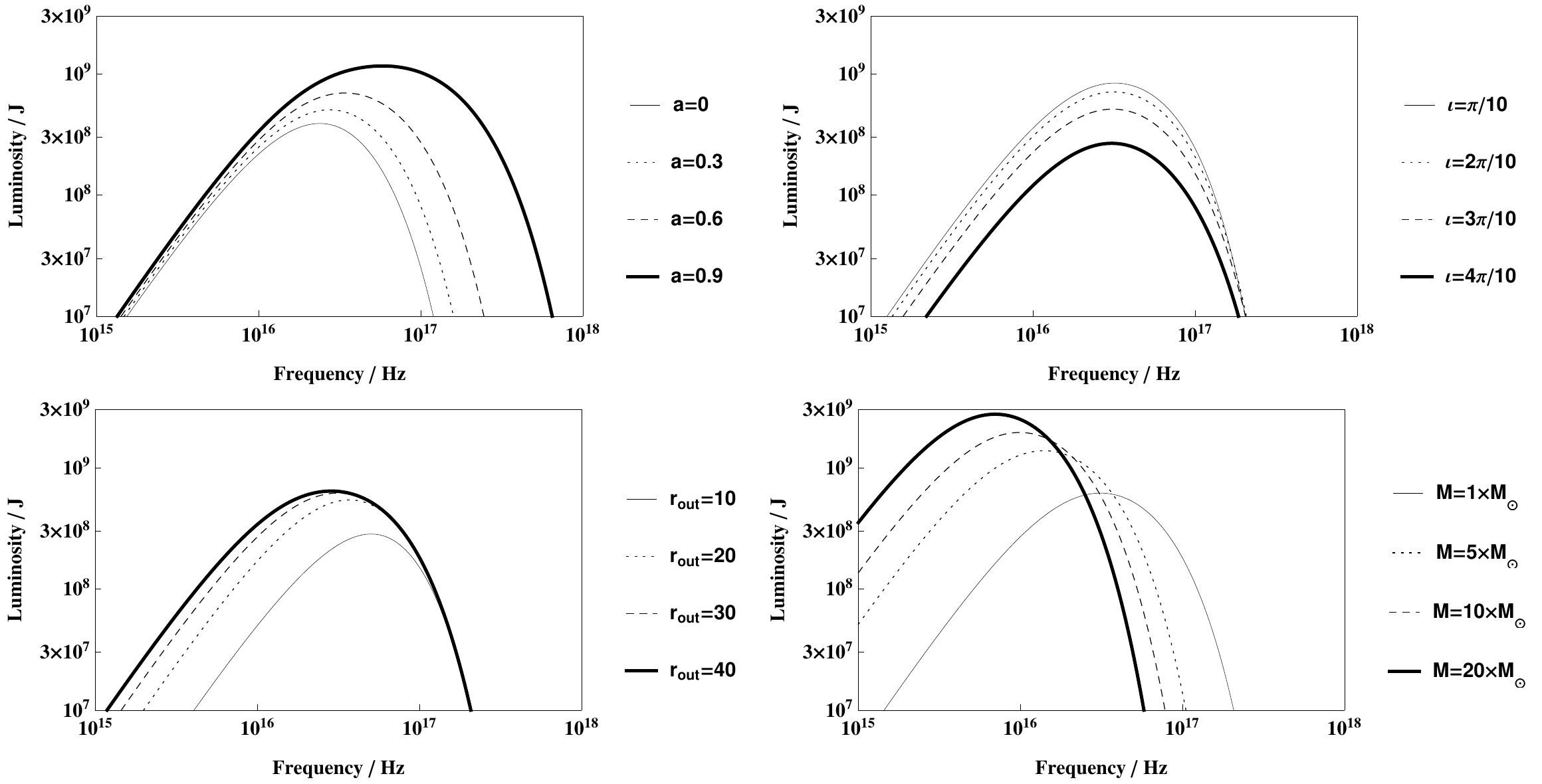}
 \caption{The dependence of the Kerr black body spectrum on various parameters. Unless indicated otherwise all parameters were set to their fiducial values, $a=0.5$, $\iota=\pi/4$, $M=\Msun$, $\dot{M}_{0}=10^{-12}\Msun/\textrm{yr}$ and $r_{\textrm{out}}=30$.}
 \label{fig:KerrTherm}
\end{figure*}

The profile is also strongly sensitive to the radius of the inner edge of the disk, which in the Kerr case (under the modelling assumptions made here) is monotonically related to the spin parameter, $a$ (see Eq.\ \ref{eq:KerrISCO}). For higher spins (and smaller values of $r_{isco}$) it can be seen from Fig.\ \ref{fig:KerrLine} that the red-shifted wing of the line profile is much more prominent. Under the assumptions that the black hole is Kerr and the inner edge of the disk is located at the ISCO the line profile can be used as an accurate probe of the black hole spin, indeed such observations are among the best evidence for the existence of near maximally spinning black holes in some active galactic nuclei (see, for example, \cite{1996MNRAS.279..837I}). 

The line profile also depends strongly on $q$; this parameter must be fit for simultaneously with all other parameters. The fiducial value $q=3$ was chosen as this is approximately the form of the gravitational energy release per unit area of the disk (see Fig.\ \ref{fig:KerrFlux}). Likewise the value of $r_{out}$ must also be fit for, but this is less problematic because (at least for $r_{out}\gtrsim 20$) the profile only depends weakly $r_{out}$.

\subsection{Black-body spectra}\label{subsec:therm}
The line emission profiles calculated above encode how the gravitational and Doppler shifting (and lightbending if included) affect a single frequency source in the disk. If instead there is a broadband source the resulting spectrum may be obtained by convolving the spectrum in the source's rest frame with the line profile obtained in the preceding section. This method can be used to calculate the black body spectra by convolving with the Plank distribution of a black body; this approach is described by \cite{1975ApJ...202..788C}.

Here we take the alternative, and more physically motivated approach of directly integrating the flux over the disk to find the spectra. From Eq.\ \ref{eq:fluxtemp4} the radial flux (shown in Fig.\ \ref{fig:KerrFlux}) is known, and assuming that the radiation is that of a black body the Stefan-Boltzmann law gives the radial temperature distribution in the disk
\begin{equation} T(r)=\left( \frac{F(r)}{\sigma} \right)^{1/4} \; ,\end{equation}
where $\sigma$ is the Stefan-Boltzmann constant. Hence the thermal spectrum may be written as the following integral where the gravitational effects on the spectrum are included through the redshift factor,
\begin{equation}\label{eq:intforthermem}F_{0}(E_{0})=\frac{8\cos \iota}{\pi}\int_{r_{isco}}^{r_{out}}\int_{0}^{2\pi}r\textrm{d}\phi\,\textrm{d}r\;\frac{E_{0}^{3}}{g^{3}\left(e^{\frac{E_{0}}{gT}}-1\right)}\; .\end{equation}
As was the case for the line emission in Eq.\ \ref{eq:fluxtemp4}, the thermal spectrum in Eq.\ \ref{eq:intforthermem} is now in the form of an integral over the disk in $r$ and $\phi$ coordinates; this was evaluated numerically. Several example thermal spectra are shown for the Kerr metric in Fig.\ \ref{fig:KerrTherm}.

The thermal spectra of the disk is the convolution between the line emission spectra and the Planck distribution. Since the Planck distribution contains no information about the gravitational field the thermal spectra in Fig.\ \ref{fig:KerrTherm} contain the same information about the black hole as the line spectra in Fig.\ \ref{fig:KerrLine}, only significantly smoothed out. Because of this smoothing it appears that the thermal spectra are less distinctive, and that therefore it would be harder to measure the parameters of the black hole using thermal emission than line emission. However, the thermal spectra may be observed across a much wider frequency range and at a larger signal-to-noise ratio. Therefore it is not obvious \emph{a priori} which method offers the best opportunity to constrain the metrics in Sec.\ \ref{sec:spacetimes}. In practice the most suitable technique depends upon the mass of the black hole; Iron line emission is typically most suitable for supermassive black holes whilst thermal emission is used for the hotter disks around stellar mass black holes. In rare instances both techniques may be used simultaneosuly, and they have been demonstrated to give consistent resuts \cite{2011MNRAS.416..941S}. Henceforth we consider only the Iron line emission, and we leave a detailed study of parameter estimation using thermal emission to future work. 

\subsection{The effect of lightbending}\label{subsec:lightbendingcalcs}
The formalism for calculating both the Iron line and thermal emission outlined in the preceding sections assumed that points in the image plane could be related to points in the disk by \emph{straight lines}, in the sense that the Boyer-Lindquist-like coordinates were treated as if they were spherical polar coordinates in flat space (these assumptions are summarised in Eq.\ \ref{eq:negligiblelightbenging}). As the light originates from the strong gravitational field the effects of lightbending (including frame dragging, if $a\neq 0$) may be significant. 

On the other hand it may be hoped that the effect of lightbending will vary slowly with changing system parameters, $\vec{\theta}$. If this is the case then while the inclusion of lightbending will have a significant impact on the observed profile it will have only a limited effect on our ability to measure the disk parameters, and hence on our ability to constrain deviations from the Kerr metric. In the language of the Fisher matrix calculations in Sec.\ \ref{sec:analysis}, even if the spectra depends strongly on whether or not lightbending is included if the derivatives of the spectra do not then the fisher matrix remains unchanged. In order to test the validity of this assumption a small number of calculations were performed including the effect of lightbending, and the results compared.

\begin{figure*}[t]
 \centering
 \includegraphics[trim=0cm 0cm 0cm 0cm, width=0.85\textwidth]{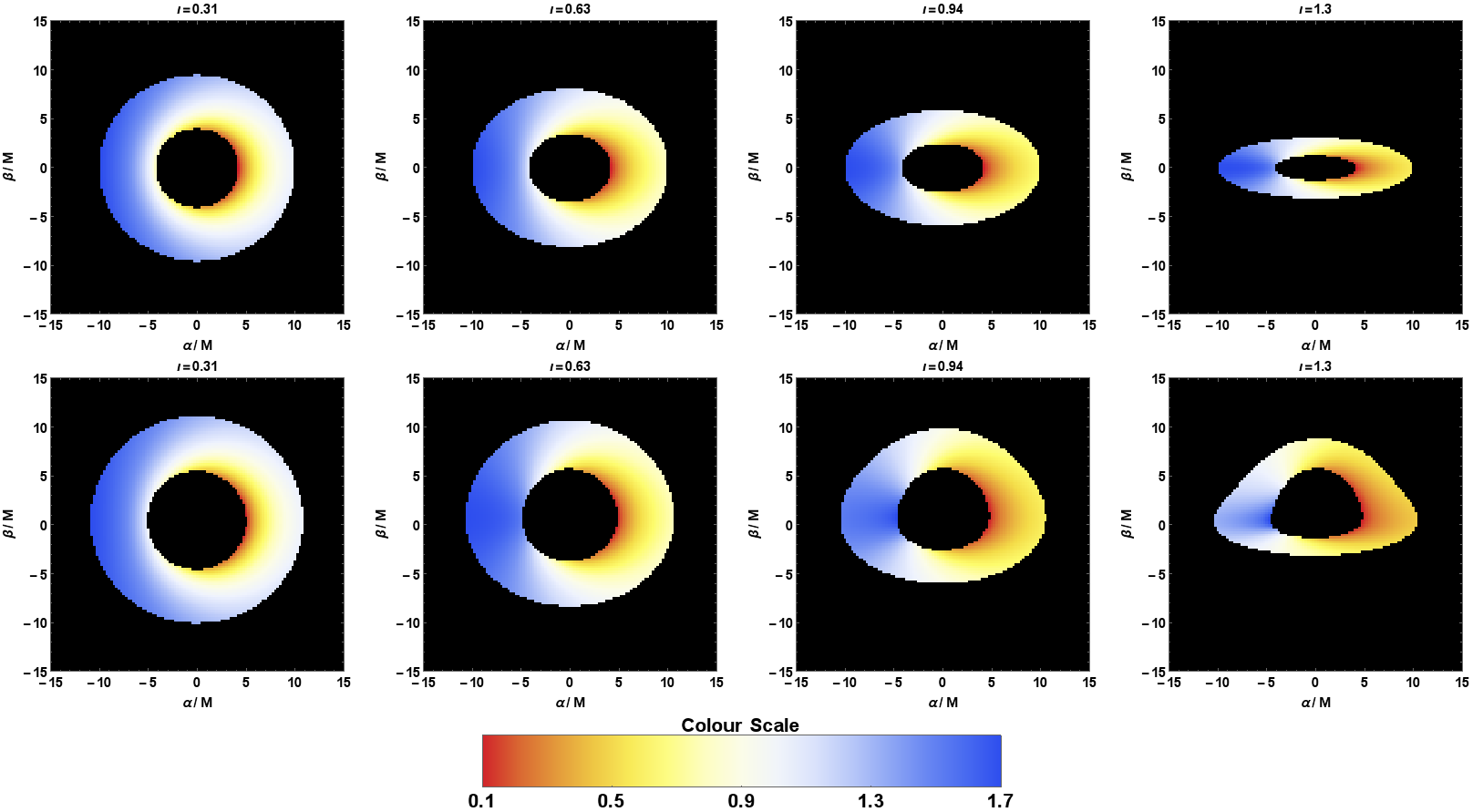}
 \caption{A set of images showing the resolved appearance of an accretion disk around a Kerr black hole. The colour indicates the redshift of the light from that portion of the disk; the top row shows the disk with no lightbending and the bottom row shows the disk including lightbending. The inclination angle is varried between plot; from left to right it takes the values $\iota =\left\{\pi /10 , 2\pi /10 , 3\pi /10 , 4\pi /10 \right\}$. The spin parameter and outer radius of the disk were fixed to $a=0.5$ and $r_{out}=10$.}
 \label{fig:LightBendingDisk}
\end{figure*}

A method for ray-tracing photons from the image plane to the disk was outlined in \cite{2012ApJ...745....1P}. The method uses the fact that the spacetime is stationary and axisymmetric to write the $t$ and $\phi$ geodesic equations for the photon in first order form, but integrates the $r$ and $\theta$ geodesic equations in second order form. This is ideal for our present purpose as the method does not require the existence of a Carter-like constant, which does not exist in all the metrics discussed in Sec.\ \ref{sec:spacetimes}. If a fourth and final constant of motion does exist (as is the case for the ${\cal{B}}_{N}$ metrics) then evaluating it along the resulting trajectory provides a convenient check on the numerical accuracy of the integration. 

The effect of lightbending on the appearance of a spatially resolved disk is shown in Fig.\ \ref{fig:LightBendingDisk} for varying inclination (the colour scale indicates redshift factor, $g$).  Particularly at high inclinations the inclusion of lightbending significantly alters the appearance of the disk; the effect is most pronounced for large values of spin and inclination where the light from the far-side, inner edge passes very close to the horizon. However, the disk is not spatially resolved by our telescope, instead we observed the integrated flux across the disk. This shows a much smaller difference. The effect of including lightbending is also more significant for higher values of the spin parameter, because the disk extends closer to the black hole where the gravitational effects are stronger.

For example, for the fiducial black hole parameters the Fisher matrix estimates an error on the dimensionless spin parameter of $\Delta a=0.06$ if light bending is neglected (see right hand panel of Fig.\ \ref{fig:FishMCMC}). If instead lightbending is included then the same Fisher matrix analysis yields an error estimate for the spin parameter of $\Delta a=0.03$. Changes in the error estimates of a factor of $\sim 2$ were observed for the other parameters. 

As anticipated above larger differences were obtained for higher values of spin and lower values of inclination, and smaller differences in the opposite limits. These changes are not enough to affect our conclusions in this paper.

\section{The Fisher matrix}\label{sec:analysis}
For each black hole metric described in Sec.\ \ref{sec:spacetimes} the Iron line and thermal spectra are characterised by a small number of parameters $\vec{\theta}$; e.g.\ for the Iron line spectra $h(\vec{\theta})$, $\vec{\theta}=(A,a,\iota,q,r_{out},\vec{\epsilon})$, where $A$ is an overall multiplicative factor relating to the unknown total luminosity and distance to the source, $a$ is the spin parameter, $\iota$ is the disk inclination, $q$ is the emissivity index described in Sec.\ \ref{subsec:line}, $r_{out}$ is the outer radius of the disk and $\vec{\epsilon}$ is the vector of any metric deformation parameters. Given a measured spectrum, $s$, the challenge is to infer the posterior probability density on these parameters, $P(\vec{\theta}|s,I)$. The peak of this distribution is positioned at the best estimate of $\vec{\theta}$, and the characteristic width of the peak in each parameter direction indicates the uncertainty. The posterior probability density is related to the likelihood of the data given the parameters, $P(s|\vec{\theta},I)$, via Bayes theorem,
\begin{equation}\label{eq:Bayes} P(\vec{\theta}|s)= \frac{P(s|\vec{\theta})P(\vec{\theta})}{P(s)}\; ;\end{equation}
where $P(\vec{\theta})$ is the prior probability density of the parameters and $P(s)$ is the normalisation constant known as the evidence,
\begin{equation} P(s)=\int\textrm{d}\vec{\theta}\;P(s|\vec{\theta})P(\vec{\theta}) \;.\end{equation}
For all calculations performed in this paper flat priors were assumed over all physically allowed regions of parameter space so the posterior is simply proportional to the likelihood within this region.

\begin{figure*}[t]
 \centering
 \includegraphics[trim=0cm 0cm 0cm 0cm, width=0.98\textwidth]{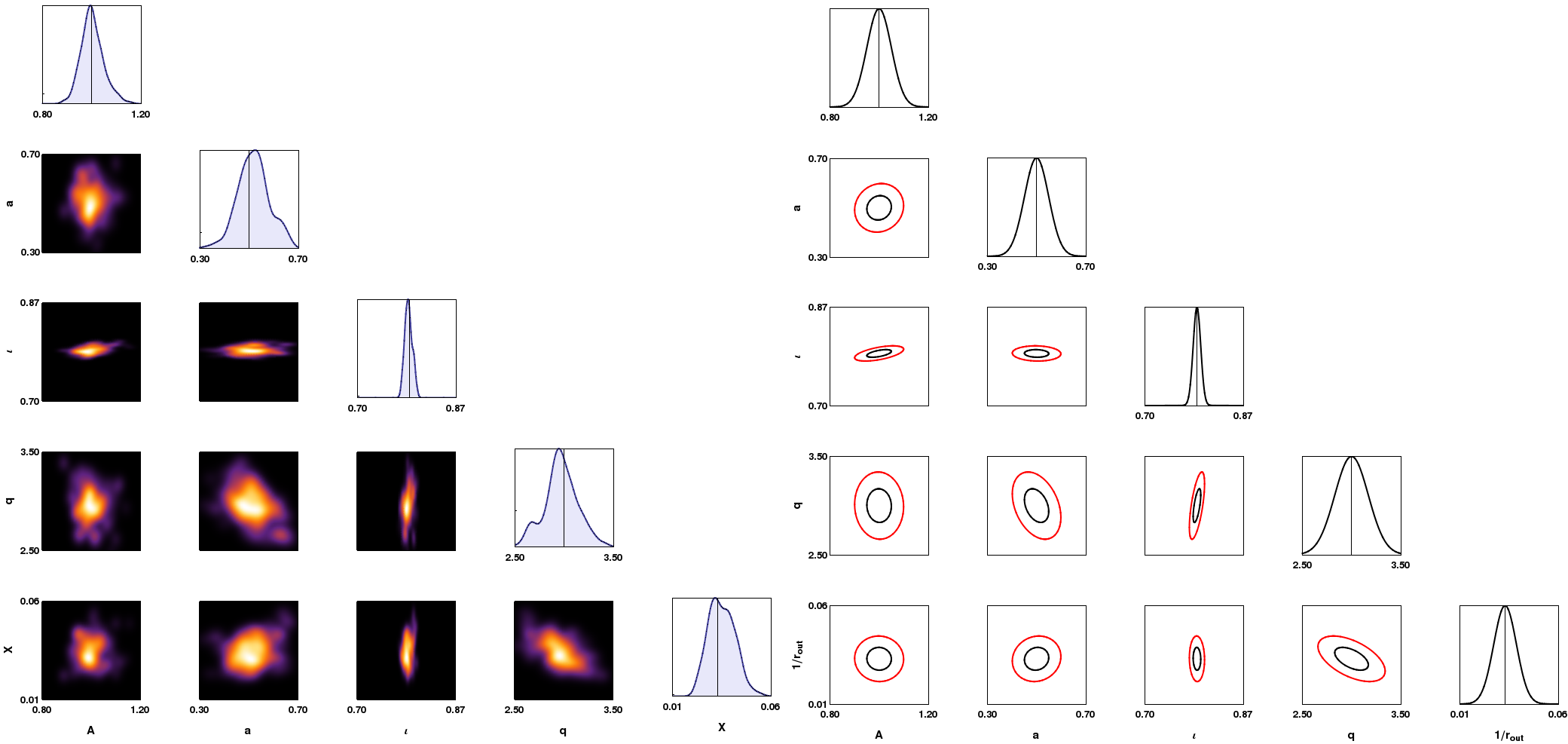}
 \caption{A comparison of the likelihood surface predicted by the Fisher matrix (indicated by $1\sigma$ (black) and $2\sigma$ (red) contours) and the true posterior (shown as a density plot using samples produced from an MCMC algorithm). All parameters were set to their fiducial values; $a=0.5$, $\iota=\pi/4$, $q=3$ and $r_{\textrm{out}}=30$.}
 \label{fig:FishMCMC}
\end{figure*}

In order to calculate the likelihood it is first necessary to make some assumptions about the performance of our detector. For simplicity we assume that across the entire energy (frequency) range the detector has a constant energy (frequency) resolution $\Delta E$ ($\Delta f$),  and that in each energy (frequency) bin there is an independent Gaussian error of size $\sigma$. The values of $\Delta E = 100\,\textrm{eV}$ ($\Delta f=2.4\times 10^{14}\,\textrm{Hz}$) and $\sigma=0.05$ were choosen to make the hypothetical instrument broadly equivalent to the best current observations. For instance, the error on the spin measurement for the fiducial disk parameters under these assumption is $\Delta a=0.06$ (see, for example, the results in Fig.\ \ref{fig:FishMCMC}).

These simplifying assumptions neglect some important effects. For example, the frequency resolution of most X-ray detectors changes substantially across observable bandwidth, the random errors in each frequency bin are due to variations in the arrival rate of photons which is a Poisson not a Gaussian process (although in the limit of high signal-to-noise the assumption of Gaussian errors becomes correct), and in addition to the uncorrelated random errors there will be systematic errors which may be correlated between frequency bins. Nevertheless, under these simplifying assumptions the likelihood is given by the following expression,
\begin{eqnarray}\label{eq:likelihood} 
&{\cal{L}}(\vec{\theta})&=\frac{\exp\left(\frac{-1}{2}\sum_{i=1}^{N}\frac{\left(s_{i}-h_{i}(\vec{\theta})\right)^{2}}{\sigma^{2}}\right)}{\sqrt{2\pi}\sigma^{N}} \nonumber \\
&&=\frac{\exp\left(\frac{-1}{2}\left<s-h(\vec{\theta})|s-h(\vec{\theta})\right>\right)}{\sqrt{2\pi}\sigma^{N}}\, ,\end{eqnarray}
where the inner product has been defined as
\begin{equation} \left<a|b\right>=\sum_{i=1}^{N}\frac{a_{i}b_{i}}{\sigma^{2}} \; . \end{equation}
In general the likelihood (Eq.~\ref{eq:likelihood}) is a complicated function of the parameters, $\vec{\theta}$. However, expanding the signal about the true parameter values, $\vec{\theta_{0}}$, (using the Einstein summation convention)
\begin{equation}\label{eq:LSA} h(\vec{\theta})=h(\vec{\theta_{0}})+\frac{\partial h}{\partial\theta_{i}}  \Big|_{\vec{\theta}=\vec{\theta}_{0}}  \delta\theta_{i}+{\cal{O}}\left((\delta\theta_{i})^{2}\right)\;,\end{equation}
and using $s=n+h(\vec{\theta_{0}})$ (where $n$ is the particular realisation of the noise observed in the detector) gives
\begin{eqnarray}\label{eq:likelihood2} 
&{\cal{L}}(\vec{\theta})&\approx \frac{\exp\left(\frac{-1}{2}\left[(n|n)-2\left(n|\frac{\partial h}{\partial\theta_{i}}\big|_{\vec{\theta}=\vec{\theta}_{0}}\delta\theta_{i}\right)+\Sigma_{ij}\delta\theta_{i}\delta\theta_{j}\right]\right)}{\sqrt{2\pi}\sigma^{N}}\, ,\nonumber\\
&&\;\textrm{where}\; \Sigma_{ij}=\left<\left.\frac{\partial h(\vec{\theta})}{\partial \theta_{i}}\Big|_{\vec{\theta}=\vec{\theta}_{0}}\right|\frac{\partial h(\vec{\theta})}{\partial \theta_{j}}\Big|_{\vec{\theta}=\vec{\theta}_{0}}\right> \; .\end{eqnarray}
Therefore, within the linear signal approximation used in Eq.\ \ref{eq:LSA}, the likelihood is a multivariate Gaussian, peaked a noise-realisation-dependent distance away from the true parameters, and with a covariance matrix given by the inverse of the Fisher information matrix, $\Sigma_{ij}$. An estimate for the error in each parameter may be read off from the corresponding component of the covariance matrix,
\begin{equation} \Delta\theta_{i}=\sqrt{\left(\Sigma^{-1}\right)_{ii}} \quad\textrm{(no sum on $i$).} \end{equation}
The Fisher matrix formalism was used first to estimate the parameter estimation accuracy for a disk in the Kerr metric (Sec.\ \ref{subsubsec:kerrres} below) and subsequently for all of the bumpy black hole spacetimes discussed in Sec.\ \ref{sec:spacetimes} (Secs.\ \ref{subsubsec:KSres} to \ref{subsubsec:B2res} below). In the case of the bumpy black hole spacetimes the true value of the bump parameters were set to zero, i.e.\ the Kerr metric was used, and the error estimate obtained for the bump is reported. The error on the bump parameter(s) are then interpreted as an estimate of the bound it may be possible to place on the size of the bump; i.e.\ if the true value of the bump parameter took this value then it would be marginally detectable with these observations. This bound should be interpreted as a lower limit, i.e.\ a best case scenario; in reality even if a non-zero value of a given deformation parameter was returned in a particular experiment it would still be a non-trivial task to rule out more mundane explanations. For example, before claiming a detection of a deviation from the Kerr solution it would presumably be necessary to consider more complicated forms for the radial emissivity, $\epsilon(r)$, than a simple power law. The free parameters in this new emissivity law would then have to be marginalised over and this would have the effect of increasing the errors on the other parameters. Other possibilities must also be considered; for example thick accretion disks, emitting material within the ISCO, reprocessing of the light by surrounding material, etc.

The applicability of the Fisher matrix rests on the validity of the linear signal approximation in Eq.\ \ref{eq:LSA}; this must hold at least within a few standard deviations from the peak in all directions in parameter space. In general it is impossible to know from the Fisher matrix alone whether one is within the region where the linear signal approximation may be safely applied. This question of the applicability of the Fisher matrix is addressed in Sec.\ \ref{subsec:validityFish}.

\section{Verification of the applicability of the Fisher matrix formalism}\label{subsec:validityFish}
The gold standard for parameter estimation is to numerically calculate the likelihood (or in general the posterior) surface over the region of parameter space of interest. This may be achieved in low dimensional problems by simply evaluating the likelihood function on a grid of parameter points. Alternatively, and more efficiently in high dimensional problems, there exist a variety of Markov chain Monte Carlo (MCMC) algorithms designed to sample points from the target probability distribution; the simplest of these is the Metropolis-Hastings algorithm. In Secs.\ \ref{subsubsec:MCMC} a MCMC analysis is performed on a sample Iron line spectra for a typical case; the resulting likelihood surfaces are compared with those predicted using the Fisher matrix.

\begin{figure*}[t]
 \centering
 \includegraphics[trim=0cm 0cm 0cm 0cm, width=0.95\textwidth]{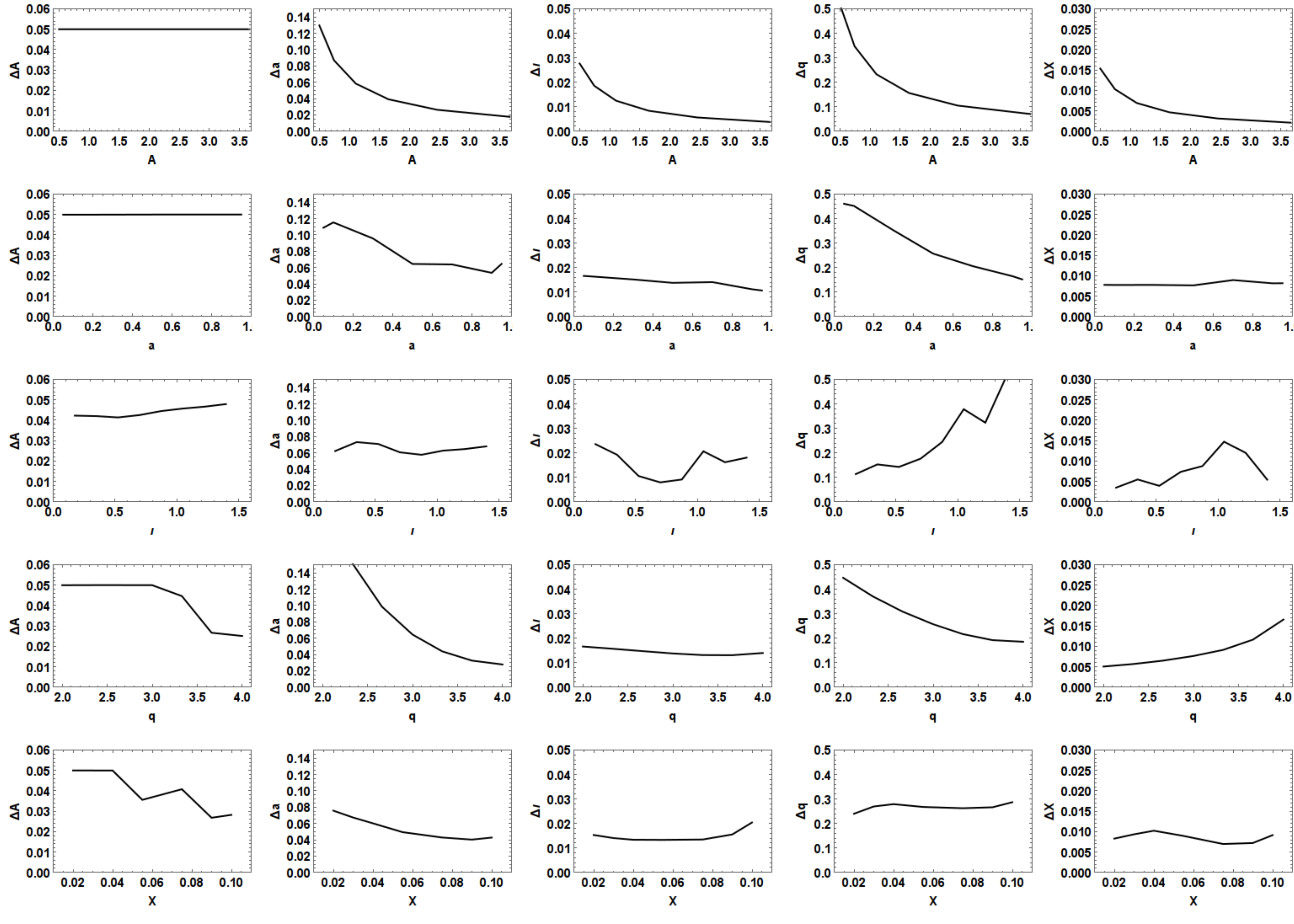}
 \caption{Each panel of this figure gives the error in the column parameter versus the value of the row parameter for fixed fiducial values of the other parameters; $a=1$, $\iota=\pi/4$, $q=3$ and $r_{\textrm{out}}=30$.}
 \label{fig:KerrFisherMatrixErrors}
\end{figure*}

MCMC type algorithms quickly become (prohibitivly) expensive as the dimension of the parameter space increases. A consistancy check on the Fisher matrix was proposed by \cite{Vallisneri} which is relativly quick to implement. The consistancy check involves testing the validity of the linear signal approximation. For some central parameter values, $\vec{\theta_{0}}$, a point $\vec{\theta}$ is picked at random from the $1\sigma$ surface estimated by the Fisher matrix. The likelihood is then evaluated exactly, and approximated with the linear signal approximation, the ratio of the two likelihoods is denoted $r(\vec{\theta})$. The logarithm of this ratio is given by
\begin{eqnarray}\label{eq:val} 
&&\left| \log   r(\vec{\theta}) \right| =\frac{1}{2}\left.\left(\Delta\theta_{i}h_{i}-\Delta h(\vec{\theta})\right\vert\Delta\theta_{i}h_{i}-\Delta h(\vec{\theta})\right) \, ,\nonumber\\
&&\quad\quad\textrm{where }\Delta\vec{\theta}=\vec{\theta}-\vec{\theta}_{0}\,,\; \Delta h (\vec{\theta})=h(\vec{\theta})-h(\vec{\theta}_{0})\,,\nonumber \\
&&\quad\quad\textrm{and }\, h_{i}=\left.\frac{\partial h(\vec{\theta})}{\partial \theta^{i}}\right|_{\vec{\theta}=\vec{\theta}_{0}} .\end{eqnarray}
Small values of $|\log r(\vec{\theta})|$ incate that the linear signal approximation is holding out as far as the $1\sigma$ surface in that particular parameter direction. This procedure may then be repeated for many points drawn randomly from the $1\sigma$ surface to assess whether the approximation holds in all directions. It should be stressed that this only checks the internal consistency of the linear signal approximation and does not guarantee the accuracy of the Fisher matrix. In Sec.\ \ref{subsubsec:val} this consistency check was performed for the Iron line emission likelihood surface, as the consistency check is faster than a full MCMC it was performed for a range of spin and inclination parameter values.

These two checks on the applicability of the Fisher matrix give us increased confidence in the results for the bounds on the various metric deformations found in Sec.~\ref{sec:results}.

\subsection{MCMC}\label{subsubsec:MCMC}
A simple Metropolis-Hastings MCMC was used to sample from the likelihood distribution in Eq.\ \ref{eq:likelihood} with the Iron line spectra described in Sec.\ \ref{subsec:line}. The resulting chain, plotted as a density histogram, is shown in Fig.\ \ref{fig:FishMCMC} alongside the Gaussian contours from the Fisher matrix analysis. From Fig.\ \ref{fig:FishMCMC} it can be seen that there is unquestionably additional structure in the true posterior which is not captured by the Fisher matrix; however the widths of the Fisher matrix Gaussians give a good indiction of the scale of the true posterior.

\begin{figure*}[t]
 \centering
 \includegraphics[trim=0cm 0cm 0cm 0cm, width=0.9\textwidth]{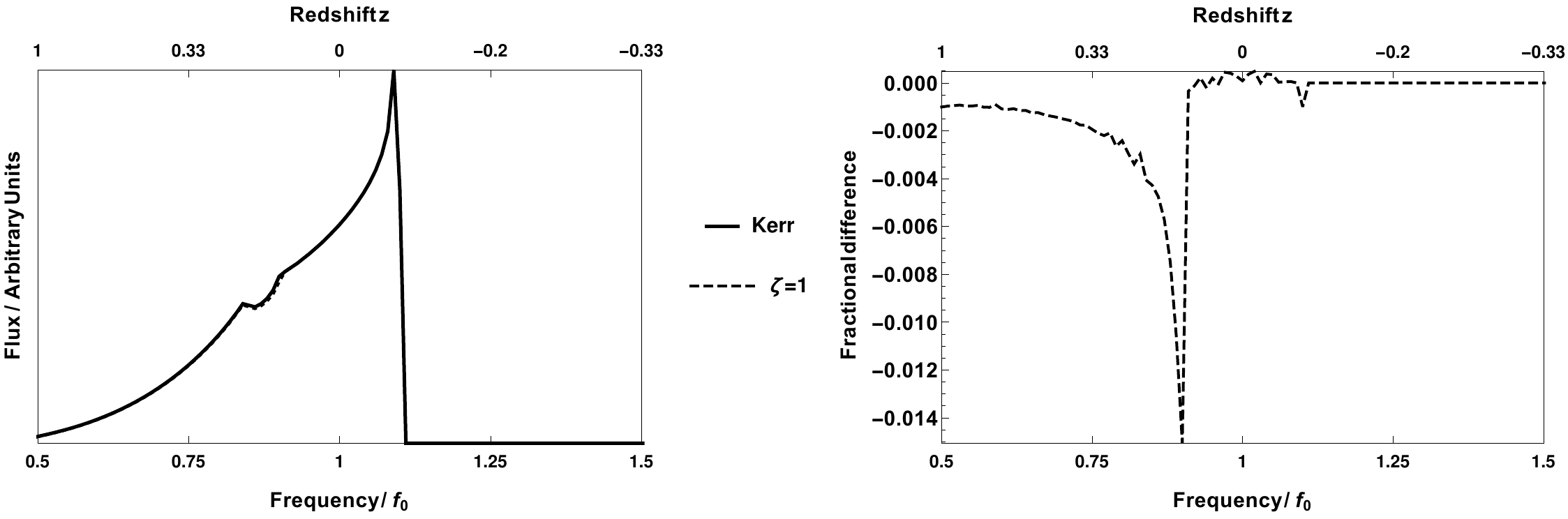}
 \caption{The left-hand panel shows a pair of Iron line spectra for accretion disks around a Kerr black hole and a quadratic in spin CS black hole (CS2 metric) with a large deformation, $\zeta=1$. Also shown in the right-hand panel is the residual plot. All parameters were set to their fiducial values; $a=0.5$, $\iota=\pi/4$, $q=3$ and $r_{\textrm{out}}=30$.}
 \label{fig:CS1IronLines}
\end{figure*}

\begin{table}[h]
\begin{center}
\begin{tabular}{ l | c c c c }
	&$\iota=\frac{\pi}{10}$&$\iota=\frac{2\pi}{10}$&$\iota=\frac{3\pi}{10}$&$\iota=\frac{4\pi}{10}$\\
\hline
$a=0.1$ & 0.0044/0.013	& 0.0088/0.032	& 0.025/0.15	& 0.10/0.50	\\
$a=0.3$ & 0.0046/0.016	& 0.0091/0.030	& 0.022/0.064	& 0.032/0.14	\\
$a=0.5$ & 0.0037/0.0069	& 0.011/0.026	& 0.019/0.083	& 0.028/0.077	\\
$a=0.7$ & 0.0039/0.0095	& 0.012/0.036	& 0.017/0.061	& 0.028/0.088	\\
$a=0.9$ & 0.0078/0.015	& 0.014/0.028	& 0.027/0.053	& 0.031/0.10	\\
\end{tabular}
\end{center}
\caption{All entries are of the form $\mu$/$M$, where $\mu$ is the mean value of the logarithm of mismatch ratio calculated on 100 points selected uniformly from the $1\sigma$ surface and $M$ is the maximum value of the mismatch ratio on the same set of points. All other parameters were set to their fiducial values of $q=3$ and $r_{\textrm{out}}=30$. (The forward slash between entries does \emph{not} indicate division.)}
\label{tab:val}
\end{table}

\subsection{Mismatch ratio}\label{subsubsec:val}
For a range of values of $a$ and $\iota$ the Fisher matrix was evaluated, and used to choose $100$ points, randomly distributed, on the $1\sigma$ surface. The mismatch ratio was evaluated for all of these points and the results are summarised in Tab.\ \ref{tab:val}; all values are less that unity, indicating that the Fisher matrix performs well for this problem.

\section{Results}\label{sec:results}
\subsection{Kerr errors: Iron line}\label{subsubsec:kerrres}
Before attempting calculations in the bumpy black hole spacetimes the Fisher matrix formalism was used to assess the accuracy with which it is possible to measure the standard parameters $\left\{A,a,\iota,q,X\right\}$ with the Iron line observations described in Sec.\ \ref{subsec:line}. The results are shown in Fig.\ \ref{fig:KerrFisherMatrixErrors}. In particular we note that the spin parameter can usually be measured with an error of $\Delta a\sim 0.1$.

\subsection{KS metric}\label{subsubsec:KSres}
\begin{figure}[h]
 \centering
 \includegraphics[trim=0cm 0cm 0cm 0cm, width=0.45\textwidth]{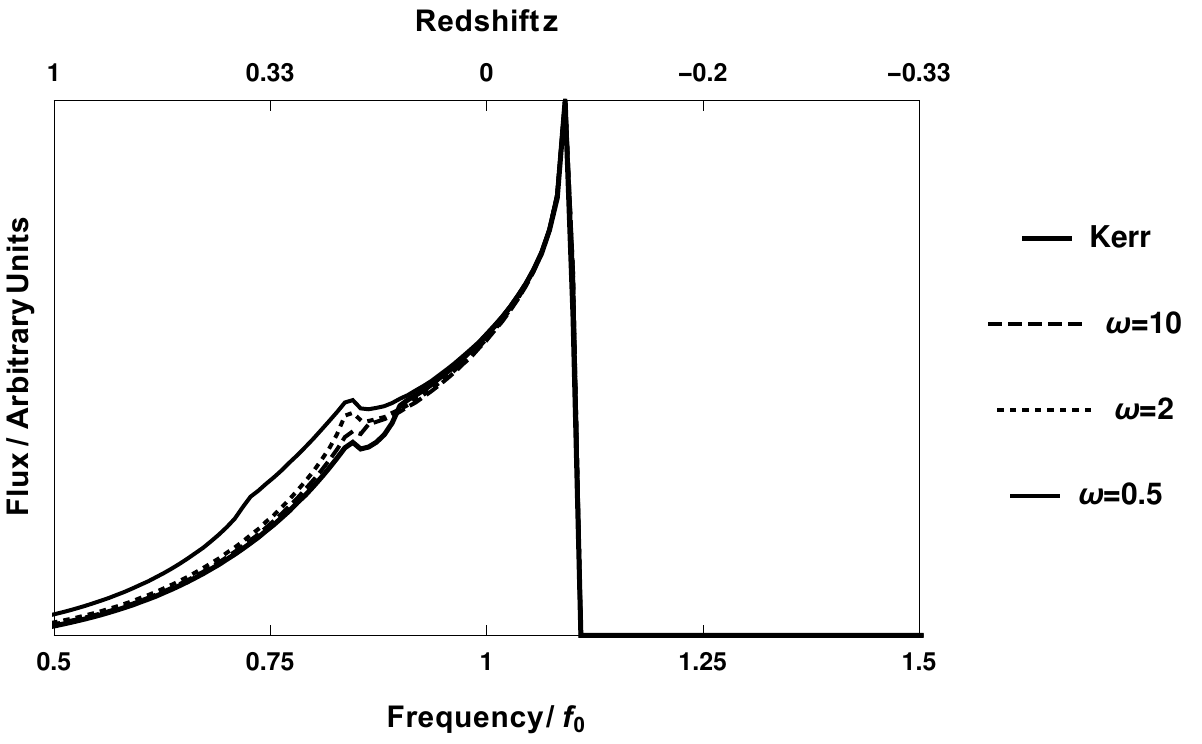}
 \caption{A series of Iron line spectra for accretion disks around KS black holes varying the value of $\omega$. All other parameters were set to fiducial values, $a=0.5$, $\iota=\pi/4$, $q=3$ and $r_{\textrm{out}}=30$.}
 \label{fig:KSIronLines}
\end{figure}

The formalism described in Sec.\ \ref{sec:emission} was used to calculate the Iron line profile in the KS metric, Eq.\ \ref{eq:KSmetric}. As described in Sec.\ \ref{sec:emission} the shift in the position of the ISCO, relative to the Kerr values, to first order in the small parameter $1/\omega$ may be calculated; $r_{isco}=r_{isco}^{\textrm{Kerr}}+\Delta r_{isco}$, where 
\begin{equation} \Delta r_{isco}=\frac{-11}{36\omega}-\frac{59a}{54\sqrt{6}\omega}\, .  \end{equation}
The ISCO moves inwards for increasing deformation, therefore the effect of a large value of $1/\omega$ is to boost the redshifted red wing of the line profile relative to the blue-shifted peak, similar to the effect of increasing spin; see Fig.\ \ref{fig:KSIronLines}. 

\begin{figure*}[t]
 \centering
 \includegraphics[trim=0cm 0cm 0cm 0cm, width=0.85\textwidth]{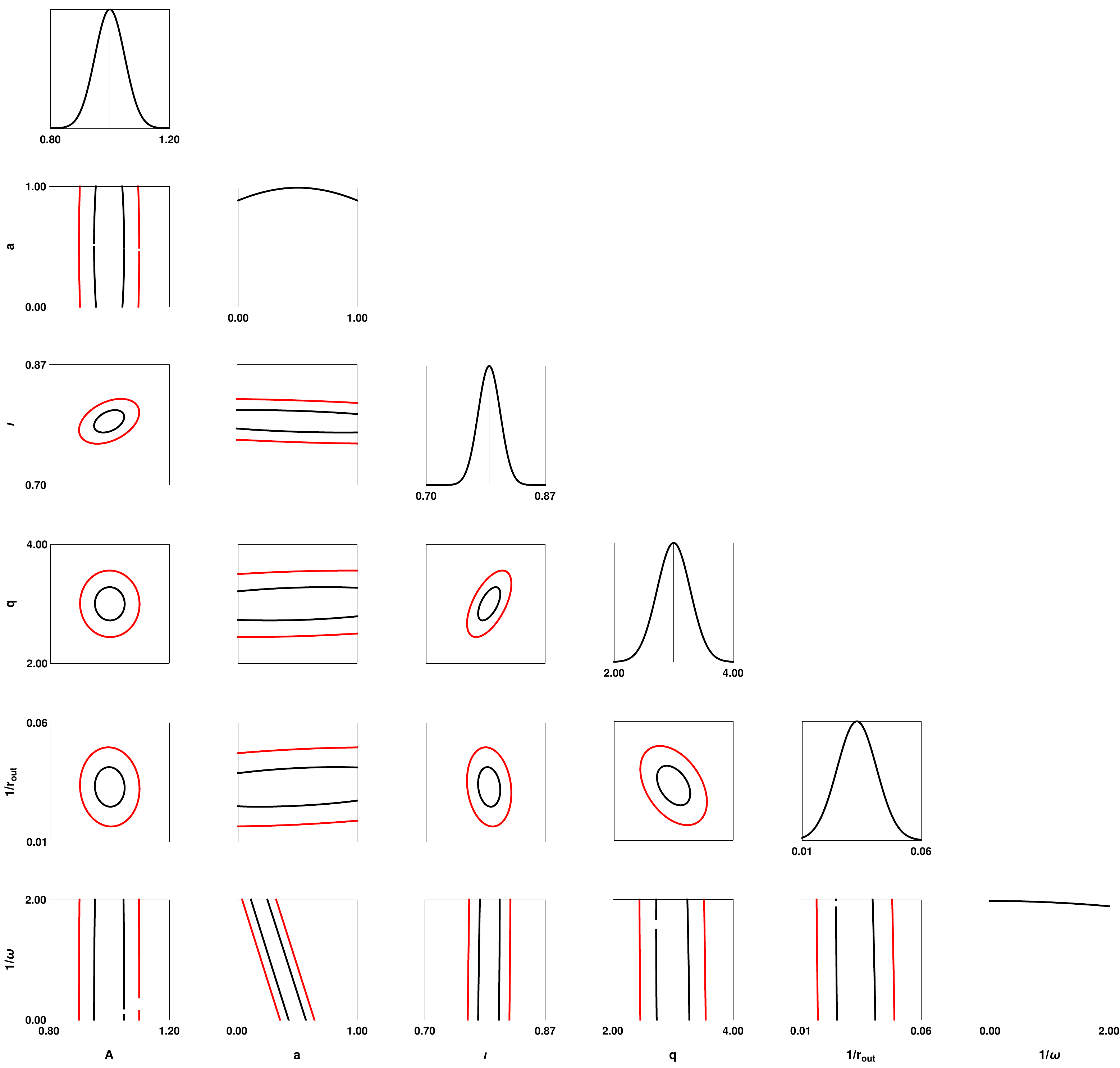}
 \caption{The $1\sigma$ and $2\sigma$ contours from a Fisher matrix analysis on the KS metric. It can be seen that introducing the extra degree of freedom in the KS deformation parameter has introduced a clear degeneracy with the spin parameter. From the bottom righthand plot it can be seen that the data is almost equally consistent with any value of $\omega$ in the range $(1/2,\infty)$, so no bound may be placed in the deformation. All parameters were set to their fiducial values; $a=0.5$, $\iota=\pi/4$, $q=3$, $r_{\textrm{out}}=30$ and $Y\equiv 1/\omega=0$.}
 \label{fig:KSFish}
\end{figure*}

The $1\sigma$ and $2\sigma$ contours from a Fisher matrix analysis are plotted in Fig.\ \ref{fig:KSFish}, it can be seen that there is a rather stark degeneracy between the spin and deformation parameter. By comparing Fig.\ \ref{fig:KSFish} to Fig.\ \ref{fig:FishMCMC} it can be seen that the errors in all other parameters are virtually unaffected by the inclusion of the deformation parameter. The effect of the degeneracy, apart from inhibiting any measurement of the spin parameter, is to make it very difficult to place any bound on the deformation; Tab.\ \ref{tab:KS} gives the bound it is possible to place on $\omega$ for different values of spin. It should be remembered that the KS metric is valid only to linear order in $a$, and in addition the metric exhibits unphysical properties (e.g. naked singularities, closed timelike curves, etc) for $\omega < 1/2$. Therefore, bearing in mind the fact that the bounds derived from the method discussed in Sec.\ \ref{sec:analysis} should be treated as lower limits, the conclusion to be drawn from Tab.\ \ref{tab:KS} is that it would be extremely difficult to place any constraint on the KS deformation parameter using observations of this type.

There have been several attempts to place bounds on the KS deformation parameter using solar system tests of gravity such as perihelion precession, deflection of light by the Sun and radar echo delay observations \citep{2011GReGr..43.1401L,2011RSPSA.467.1390H,2011IJMPD..20.1079I}. The bounds obtained from these studies are typically on the length scale $\omega\gtrsim 5\times 10^{-28}\,\textrm{m}$, corresponding to a dimensionless bound $\omega M^{2} \gtrsim 3\times 10^{-16}$, with $M=\Msun$. It should be noted that these bounds are much less stringent than the requirement imposed here that $\omega>1/2$, which is necessary to ensure that there is an event horizon. However, as these tests were conducted in the weak field around a material object whose radius is much larger than the gravitational radius the vacuum solution is not valid down near the event horizon and the constriant $\omega>1/2$ need not apply.

\begin{table}[h]
\begin{center}
\begin{tabular}{ l | c  }
	&$\Delta (1/\omega)$\\
\hline
$a=0.1$ &  15.2 \\
$a=0.3$ &  9.23 \\
$a=0.5$ &  5.47 \\
$a=0.7$ &  4.76 \\
$a=0.9$ &  3.44 \\
\end{tabular}
\end{center}
\caption{The Bounds it is possible to place on the small KS deformation parameter ($1/\omega$) for different values of the spin parameter, $a$. All other parameters were set to their fiducial values; $\iota=\pi/4$, $q=3$ and $r_{\textrm{out}}=30$ and $Y\equiv 1/\omega=0$.}
\label{tab:KS}
\end{table}

\subsection{CS metric}\label{subsubsec:CS1res}
Shown in Fig.\ \ref{fig:CS1IronLines} is the Iron line profile for a Kerr black hole and a CS2 black hole with deformation $\zeta=1$. It should be remembered that the CS1 and CS2 metrics in Eqs.\ \ref{eq:CS1metric} and \ref{subsec:CS2} are only valid in the limit $\zeta\ll 1$. However, even with the large deformation used in Fig.\ \ref{fig:CS1IronLines} the change in the line profile is only visible in the residual plot shown in the right-hand panel. The Iron line profile for the CS1 metric shows very similar behaviour to that in Fig.\ \ref{fig:CS1IronLines}.

Bearing in mind the results in Fig.\ \ref{fig:CS1IronLines}, it is clear that it will be extremely difficult to bound the CS1 deformation parameter using this technique. Tab.\ \ref{tab:CS1} shows the bounds it is possible to place on both the CS1 and CS2 deformation parameters for a range of values of $a$; it was found that no bounds less than unity were possible with Iron line observations, however, the best results were obtained for the CS1 metric and high values of spin. Both the CS1 and CS2 metrics are expansions in the $a$ parameter, therefore the bounds for the higher values of spin (particularly $a=0.9$) should be treated with more caution than the low spins. However, this does not effect our main conclusion that no bounds less than unity were possible. 

For comparison, weak field tests using the frame-dragging effect around the Earth measured by the Gravity Probe B and the LAGEOS satellites places a bound $\zeta^{1/4}<10^{8}\,\textrm{km}$ \citep{2011PhRvD..84l4033A}. Tighter bounds will be possible using strong field tests, for example it was found that bounds of $\zeta^{1/4}<10^{4}\,\textrm{km}$ corresponding to a dimensionless bound of $\zeta < \times 10^{-7}$ would be possible with eLISA observations of EMRIs \citep{2012PhRvD..86d4010C}.

\begin{table}[h]
\begin{center}
\begin{tabular}{ l | c  c }
	&$\Delta \zeta_{\textrm{CS1}}$ &$\Delta\zeta_{\textrm{CS2}}$\\
\hline
$a=0.1$ & 101  & 122	\\
$a=0.3$ & 97.0 & 62.6	\\
$a=0.5$ & 24.0 & 25.4	\\
$a=0.7$ & 23.9 & 9.98	\\
$a=0.9$ & 1.32 & 5.34	\\
\end{tabular}
\end{center}
\caption{The Bounds it is possible to place on the CS1 and CS2 deformation parameters for different values of the spin parameter, $a$. It should be remembered that the CS parameter is constrained to be $\zeta\ll 1$, therefore no meaningful constraint may be placed with these observations. All other parameters were set to fiducial values, $\iota=\pi/4$, $q=3$, $r_{\textrm{out}}=30$ and $\zeta=0$.}
\label{tab:CS1}
\end{table}

\subsection{${\cal{B}}_{N}$ metrics}\label{subsubsec:B2res}

\begin{figure*}[t]
 \centering
 \includegraphics[trim=0cm 0cm 0cm 0cm, width=0.9\textwidth]{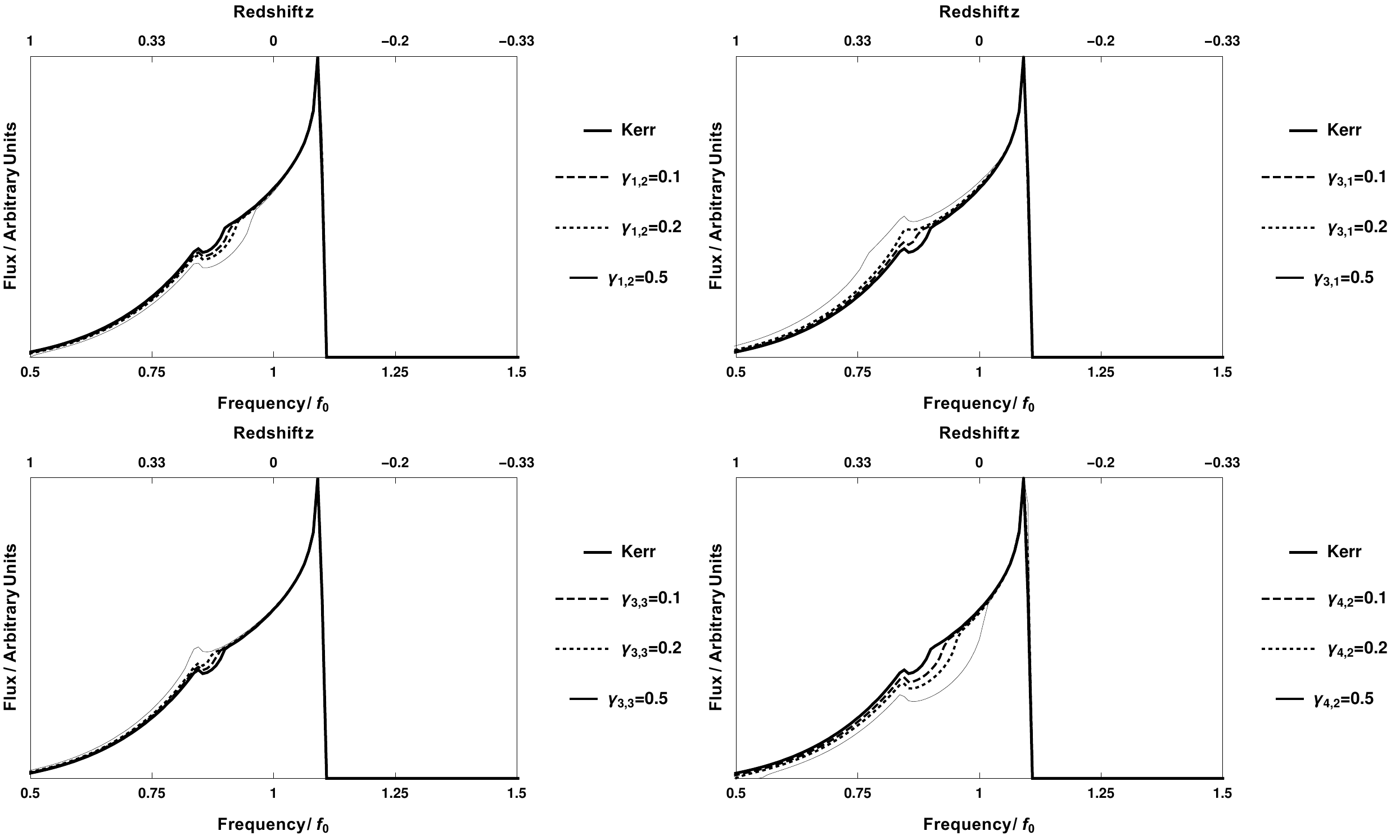}
 \caption{A series of Iron line spectra for accretion disks around B2 bumpy black holes with varying values of the bump parameters. All other parameters were set to fiducial values; $a=0.5$, $\iota=\pi/4$, $q=3$, $r_{\textrm{out}}=30$, and $\gamma_{i,j}=0$ unless otherwise indicated.}
 \label{fig:B2Line}
\end{figure*}

\begin{figure*}[t]
 \centering
 \includegraphics[trim=0cm 0cm 0cm 0cm, width=0.9\textwidth]{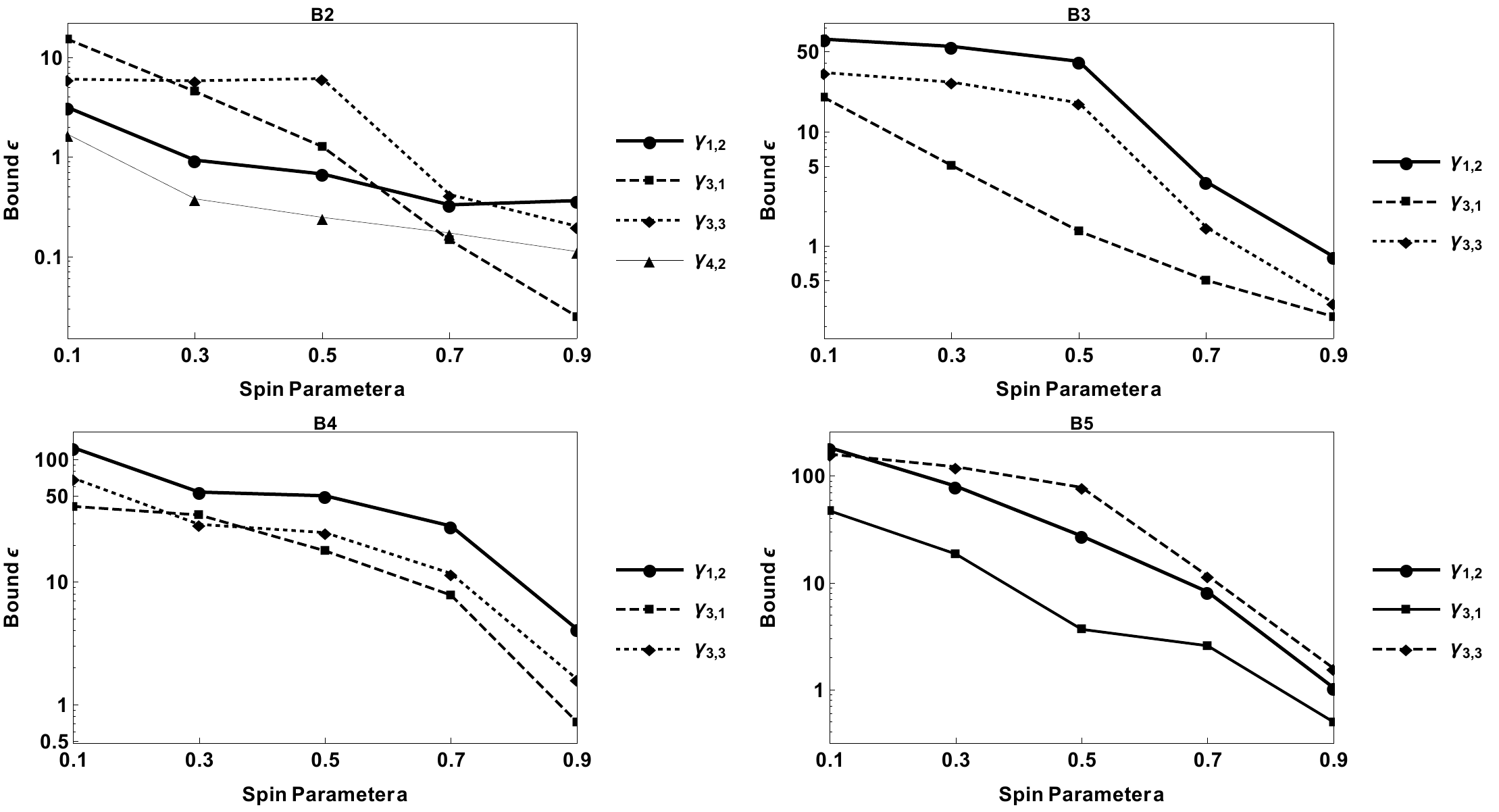}
 \caption{The bounds it is possible to place on the different ${\cal{B}}_{N}$ bump parameters for different values of the spin parameter, $a$. The tightest bounds can typically be placed for low values of $N$ (i.e.\ ${\cal{B}}_{2}$) and for the most highly spinning black holes. All other parameters were set to fiducial values; $\iota=\pi/4$, $q=3$, $r_{\textrm{out}}=30$ and $\epsilon =0$.}
 \label{fig:summary}
\end{figure*}

Shown in Fig.~\ref{fig:B2Line} are a series of Iron line profiles for the metrics defined in Sec.\ \ref{sec:spacetimes} with the constants ${\cal{B}}_{2}$ varied. The deformation parameters were varied between 0 and 0.5. Shown in Fig.~\ref{fig:summary} is the bounds it is possible to place on the different deformations using Iron line observations for different values of $a$.

The results in Fig.~\ref{fig:summary} show that the tightest constraints on the ${\cal{B}}_{N}$ bumpy black holes can be placed on the lowest values of $N$ for the highest values of spin. The lower values of $N$ have deformations entering at lower powers of $1/r$, since all the emission originates from outside the horizon (where $r>1$) it is to be expected that deformations at lower $N$ are easier to constrain. It is also to be expected that higher values of spin make placing constraints easier, because the ISCO moves to smaller values of $r$ for larger spins, and most of the emission comes from close to the ISCO, the deformation has a greater effect on the spectra for high spins.

\section{Discussion}\label{sec:discussion}
Observations of accretion disks offer an enticing opportunity to probe strong gravitational fields. Such observations have already provided some of the best evidence for the existence of black holes (in particular the existance of an event horizon, and a separate innermost stable circular orbit) and are routinely used to measure the spin of black holes. In this paper we have considered the possibility of using X-ray emission from a thin accretion disk to distinguish between the Kerr black hole predicted by GR and alternative black hole metrics. For this purpose a series of parametrically deformed Kerr metrics, referred to as bumpy black holes, expanded in powers of $M/r$, were used \cite{2011PhRvD..83j4027V,2011PhRvD..84f4016G}. In this paper both the transition line and the thermal emission around these general bumpy black holes were calculated. The Fisher matrix formalism was used for the Iron line emission to estimate the accuracy to which the various disk parameters could be measured. 

In addition to this general family of bumpy black holes the disk emission in several specific, known black hole spacetimes were also considered. In particular the Iron line emission in the metric CS2 due to \cite{2012PhRvD..86d4037Y} was considered for the first time in the literature. For the CS2 metric it was found that it is impossible to place constraints on the small, dimensionless coupling parameter less than unity.

The technique of using accretion disk observations to constrain alternative black hole solutions has limitations. Qualitatively, the method works well for measuring the spin of a Kerr black hole, but adding extra deformation parameters typically introduces degeneracies which make constraining the bump parameters very difficult. This was very clearly seen in the case of the Kehagias Sfetsos metric, see Fig.~\ref{fig:KSFish}. For the BN metrics it was found that bounds less than unity were only possible in a limited number of cases, typically for the leading order deformations around highly spinning black holes. It should also be remembered that the bounds presented here are optimistic, ``best case'' scenarios for the reasons discussed in Sec.\ \ref{sec:analysis}.

In the near future it will become possible to perform this sort of test of GR using gravitational wave observations of the inspiral and coalescence of a binary containing neutron stars or black holes. In particular, the possibility of using eLISA observations of extreme mass ratio inspirals to constrain the CS1 metric was recently considered in \cite{2012PhRvD..86d4010C}. It was found that bounds on the dimensionless deformation parameter $\epsilon<10^{-6}$ may be possible, several orders of magnitude better than the bounds estimated here. As the CS1 metric is a special case of the general ${\cal{B}}_{N}$ metrics considered here, it would be interesting to try to constrain the ${\cal{B}}_{N}$ metrics here using the same techniques and see if similar improvements are possible.

\clearpage

\appendix
\section{Metric components}\label{app:a}
\subsection{BN}\label{subsec:BNcomponents}
For the bumpy ${\cal{B}}_{N}$ metrics discussed in Sec.\ \ref{sec:spacetimes} all of the metric coefficients up to ${\cal{O}}(1/r^{5})$ are reproduced here.
\begin{eqnarray}
&h_{tt,2}=&\gamma_{1,2}+2\gamma_{4,2}-2a\gamma_{3,1}\sin^{2}\theta \label{eq:bigEqstart}\nonumber\\
&h_{tt,3}=&\gamma_{1,3}-8\gamma_{4,2}-2\gamma_{1,2}+2\gamma_{4,3}+8a\gamma_{3,1}\sin^{2}\theta \nonumber\\
&h_{tt,4}= & -8\gamma_{4,3}-2\gamma_{1,3}+2\gamma_{4,4}+8\gamma_{4,2}+\gamma_{1,4}-8a\gamma_{3,1}\sin^{2}\theta  \nonumber\\
         & & +a^{2}\left(\gamma_{1,2}+2\gamma_{4,2}\right)\sin^{2}\theta+2a^{3}\gamma_{3,1}\cos^{2}\theta \sin^{2}\theta \nonumber\\
&h_{tt,5}= & 16a^{3}\gamma_{3,1}\sin^{4}\theta + \sin^{2}\theta \left[ 4a\gamma_{3,3}+a^{2}\left(\gamma_{1,3}-2\gamma_{1,2}  \right.\right. \nonumber \\
&&\left.\left.-12\gamma_{4,2}+2\gamma_{4,3}\right)-12a^{3}\gamma_{3,1} \right] +a^{2}\left(8\gamma_{4,2}+2\gamma_{1,2}\right) \nonumber\\
&  &  + \gamma_{1,5} +2\gamma_{4,5}-2\gamma_{1,4}+8\gamma_{4,3}-8\gamma_{4,4} 
\end{eqnarray}
\begin{eqnarray}
&h_{t\phi,2}=&-M\sin^{2}\theta\left[ \gamma_{3,3}+a\left(\gamma_{1,2}+\gamma_{4,2} \right) +a^{2}\gamma_{3,1} \right] \nonumber\\
&h_{t\phi,3}=&-8Ma^{2}\gamma_{3,1}\sin^{4}\theta \nonumber\\
&&+M\sin^{2}\theta\left[ \left(2\gamma_{3,3}-\gamma_{3,4} \right)+a\left(6\gamma_{4,2}-\gamma_{4,3} \right.\right.\nonumber\\
&&\left.\left.+2\gamma_{1,2}-\gamma_{1,3} \right)+2\gamma_{3,1}a^{2} \right] \nonumber\\
&h_{t\phi,4}= & M\sin^{4}\theta\left[a^{2}\left(8\gamma_{3,1}-\gamma_{3,3} \right)+a^{3}\left(-\gamma_{1,3}-\gamma_{4,2} \right)\right.\nonumber \\
&&\left.-a^{4}\gamma_{3,1} \right] + \sin^{2}\theta \left[  \left( 2\gamma_{3,4}-\gamma_{3,5} \right)+a\left(-\gamma_{4,4}\right.\right.\nonumber\\
&&\left.\left.-8\gamma_{4,2}+6\gamma_{4,3}-\gamma_{1,4}+2\gamma_{1,3} \right)  -a^{2}\gamma_{3,3}  \right] \nonumber\\
&h_{t\phi,5}= & -16Ma^{4}\gamma_{3,1}\sin^{6}\theta\nonumber\\
&&+M\sin^{4}\theta\left[ a^{2}\left(-2\gamma_{3,3}-\gamma_{3,4}\right)\right.\nonumber\\
&&\left.+a^{3}\left(\gamma_{4,3}+10\gamma_{4,2}+2\gamma_{1,2}-\gamma_{1,3}\right) +14a^{4}\gamma_{3,1} \right] \nonumber\\
& &  +\sin^{2}\theta\left[ \left(2\gamma_{3,5}-\gamma_{3,6}\right)\right.\nonumber\\
&&\left.+a\left(-\gamma_{1,5}-8\gamma_{4,3}-\gamma_{4,5}+2\gamma_{1,4}+6\gamma_{4,4}\right) \right. \nonumber\\
& & \left. -\gamma_{3,4}a^{2}+a^{3}\left(-2\gamma_{1,2}-6\gamma_{4,2}\right)-2a^{4}\gamma_{3,1} \right]
\end{eqnarray}
\begin{eqnarray}
&h_{rr,2}=&-\gamma_{1,2}\nonumber\\
&h_{rr,3}=&-\gamma_{1,3}-2\gamma_{1,2}, \nonumber\\
&h_{rr,4}=&-\gamma_{1,4}-2\gamma_{1,3}-4\gamma_{1,2}+(1/2)\gamma_{1,2}a^{2}\left(1-\cos 2\theta \right) \nonumber\\
&h_{rr,5}=&a^{2}\sin^{2}\theta\left(\gamma_{1,3}+2\gamma_{1,2} \right)\nonumber\\
&&-\gamma_{1,5}-2\gamma_{1,4}-4\gamma_{1,3}-8\gamma_{1,2}+2a^{2}\gamma_{1,2} 
\end{eqnarray}
\begin{eqnarray}
&h_{\phi\phi,-2}=&0\nonumber\\
&h_{\phi\phi,-1}=&0\nonumber\\ 
&h_{\phi\phi,0}=&2M^{2}a\gamma_{3,1}\sin^{4}\theta\nonumber\\
&h_{\phi\phi,1}=&0 \nonumber\\
&h_{\phi\phi,2}=&M^{2}\sin^{4}\theta\left[ 2a\gamma_{3,3}+a^{2}\gamma_{1,2}+a^{3}\gamma_{3,1}\left( 4-2\cos^{2}\theta \right) \right] \nonumber\\
&h_{\phi\phi,3}= & 8M^{2}a^{3}\gamma_{3,1}\sin^{6}\theta + M^{2}\sin^{4}\theta\left[ a\left(-4\gamma_{3,3}+2\gamma_{3,4} \right)\right.\nonumber\\
&&\left.+a^{2}\left( -2\gamma_{1,2} -4\gamma_{4,2}+\gamma_{1,3} \right)-4a^{3}\gamma_{3,1} \right] \label{eq:bigEqend}
\end{eqnarray}

\subsection{CS2}\label{subsec:CS2components}
For the CS2 metric discussed in Sec.\ \ref{subsec:CS2} the metric perturbations are reproduced here, with $f(r)=1-(2/r)$.
\begin{eqnarray}
\delta\left(g_{tt}^\CSt \right) &=& \zeta a^2 \frac{1^3}{r^3} \Bigg[  \frac{201}{1792} \left( 1+\frac{1}{r} +\frac{4474}{4221} \frac{1^2}{r^2} \right.\nonumber\\
&&\left.-\frac{2060}{469} \frac{1^3}{r^3}+\frac{1500}{469} \frac{1^4}{r^4} - \frac{2140}{201} \frac{1^5}{r^5}  \right.\nonumber \\
&&\left.+ \frac{9256}{201} \frac{1^6}{r^6}- \frac{5376}{67} \frac{1^7}{r^7}  \right) (3\cos^2 \theta -1)  \nonumber \\
&&- \frac{5}{384} \frac{1^2}{r^2} \left( 1 + 100 \frac{1}{r} \right.\nonumber\\
&&\left.+ 194\frac{1^2}{r^2} + \frac{2220}{7} \frac{1^3}{r^3} - \frac{1512}{5} \frac{1^4}{r^4} \right) \Bigg]    \,, 
\\
\delta\left(g_{t\phi}^\CSt \right) &=& \frac{5}{4}\zeta\chi\frac{1}{r^{4}}\left(1+\frac{12}{7r^{2}}+\frac{27}{10r^{2}}\right)
\\
\delta\left(g_{rr}^\CSt \right) &=&   \zeta a^2 \frac{1^3}{r^3 f(r)^2} \Bigg[  \frac{201}{1792}  f(r) \left( 1+ \frac{1459}{603} \frac{1}{r} \right.\nonumber\\
&&\left. +\frac{20000}{4221} \frac{1^2}{r^2}+\frac{51580}{1407} \frac{1^3}{r^3} -\frac{7580}{201} \frac{1^4}{r^4} \right. \nonumber \\ 
& & \left. - \frac{22492}{201} \frac{1^5}{r^5}  - \frac{40320}{67} \frac{1^6}{r^6} \right) (3 \cos^2 \theta -1)   \nonumber \\
& & - \frac{25}{384} \frac{1}{r} \left( 1 + 3\frac{1}{r} + \frac{322}{5} \frac{1^2}{r^2} + \frac{198}{5} \frac{1^3}{r^3} \right.\nonumber\\
&&\left.+ \frac{6276}{175} \frac{1^4}{r^4} - \frac{17496}{25} \frac{1^5}{r^5}   \right) \Bigg]  \,,
\\
\delta\left(g_{\theta\theta}^\CSt \right) &=& \frac{201}{1792} \zeta a^2 1^2 \frac{1}{r} \left( 1 + \frac{1420}{603} \frac{1}{r} + \frac{18908}{4221} \frac{1^2}{r^2} \right. \nonumber \\
& & \left. + \frac{1480}{603} \frac{1^3}{r^3}+ \frac{22460}{1407} \frac{1^4}{r^4} \right.\nonumber\\
&&\left.+ \frac{3848}{201} \frac{1^5}{r^5} + \frac{5376}{67} \frac{1^6}{r^6} \right) (3 \cos^2 \theta -1)
\\
\delta\left(g_{\phi\phi}^\CSt \right) &=&  \sin^2 \theta g_{\theta\theta}^\CSt \; .
\end{eqnarray}

\clearpage

\section{Radial dependence of the flux}\label{app:b}
\cite{1974ApJ...191..499P} derive various expressions for the radial structure of the disk, including an expression for the radial dependence of the flux (Eq.\ \ref{eq:radialfluxmain} in the main text). Here the derivation is summarised for completeness.

The thin disk is assumed to be axisymmetric, stationary, and lying in the equatorial place; therefore all quantities in the disk depend only on the radial coordinate. It is assumed that the material in the disk moves (very nearly) on circular geodesics. The four-velocity of individual fluid elements, $u^{\mu} _{0}$, when mass averaged over the disk structure must therefore be the four-velocity of the geodesic orbit $u^{\mu}$ in Eq.\ \ref{eq:four:vel}.
\begin{equation}\label{eq:diskvelocity} u^{\mu} = \frac{1}{\Sigma(r)}\int_{-h}^{+h}\textrm{d}z\; \rho_{0} u^{\mu} _{0}  \, ,\;
\textrm{where} \;  \Sigma (r) = \int_{-h}^{+h}\textrm{d}z\;  \rho _{0} \, . \end{equation}
Where $\rho_{0}$ mass density in the rest frame of the orbiting material. Without loss of generality the stress-energy tensor may be decomposed by writing
\begin{equation}\label{eq:stresstensor} T^{\mu \nu} = \rho_{0}\left( 1+\Pi \right)u^{\mu}u^{\nu} + t^{\mu\nu} + 2u^{(\mu}q^{\nu )} \end{equation}
where the physical interpretation of each term becomes clear in the rest frame of the orbiting material; $\Pi$ is the specific internal energy, $t^{\mu\nu}$ is the stress tensor in the averaged rest frame of the material and $q^{\mu}$ is the energy flow vector. The round brackets in the superscript of Eq.\ \ref{eq:stresstensor} denote symmeterisation with respect to the enclosed indices. The tensors $t^{\mu\nu}$ and $q^{\mu}$ obey the orthogonality relations $u_{\mu}q^{\mu}=0$ and $u_{\mu}t^{\mu\nu}=u_{\nu}t^{\mu\nu}=0$. By analogy with $\Sigma(r)$ in Eq.\ \ref{eq:diskvelocity} the average stresses in the disk are defined as
\begin{equation} W_{\mu}^{\; \nu} = \int_{-h}^{+h}\textrm{d}z\; t_{\mu}^{\; \nu}  \; . \end{equation}

It is also assumed that heat flow within the disk is negligible except for in the vertical direction, which is reasonable as the disk is thin. The non-local heating effects due to light emitted by one portion of the disk being re-absorbed by another portion are also neglected.
\begin{equation}\label{eq:heatflow} q^{t}=q^{r}=q^{\phi}= 0\;,\quad \textrm{at}\quad z=\pm h.\end{equation}
Since the only time-averaged stress that reaches out of the disk to infinity is carried by photons (neglecting gravitational radiation and any coherent superposition of long wavelength radiation), and using Eq.\ \ref{eq:heatflow}, on the upper and lower edges of the disk the following terms of the stress-tensor disappear;
\begin{eqnarray}\label{eq:tcomp} &t_{\phi}^{\; z}= t_{r}^{\; z}= t_{t}^{\; z} =0 \nonumber \\
&\textrm{and}\; \left| q^{z} \right|=F(r)\,,\;\textrm{at}\; z=\pm h \,.\end{eqnarray}

Since we are assuming the particles in the disk are, very nearly, on circular, geodesic orbits it follows that the acceleration due to pressure gradients in the disk must be much less than the acceleration due to gravity otherwise the material would be pushed off its geodesic trajectory. Using the approximate relation $t_{rr}\approx\rho_{0}\Pi$ (which is valid for any astrophysically reasonable matter \cite{1974ApJ...191..499P}) leads directly to the condition of negligible specific heat;
\begin{eqnarray}
&\textrm{radial pressure acceleration} \approx \left| \partial_{r}t_{rr}/\rho_{0} \right| \approx \left| \partial_{r}\left( t_{rr}/\rho_{0}\right)  \right| \nonumber\\
&\textrm{gravitational acceleration} \approx | \partial_{r}\tilde{E} | \approx | \partial_{r}( 1-\tilde{E} ) | \nonumber\\
& | \partial_{r} (t_{rr}/\rho_{0} ) | \ll | \partial_{r}( 1-\tilde{E} ) |\quad \Rightarrow \quad \Pi \ll 1-\tilde{E}\; .
\end{eqnarray}
If the internal energy is small compared to the gravitational potential energy, this means that as the material spirals in towards the black hole all of the gravitational potential energy is radiated away.

With these simplifying assumptions in place the equations governing the structure of the disk follow from the conservation of stress-energy ($\nabla_{\mu}T^{\mu\nu}=0$), and the conservation of rest mass of the fluid \citep{MTW},
\begin{equation} \nabla_{\mu} \left( \rho_{0}u^{\mu} \right)=0  \; .\end{equation}
This is integrated over the spacetime volume $\left\{{\cal{V}}: t\in\left(t_{0},t_{0}+T\right), \right.\allowbreak\left. r\in\left(r,r+\Delta r\right), \right.\allowbreak\left. \phi\in\left(0,2\pi\right), \right.\allowbreak\left. z\in\left(-h,+h\right)\right\}$. Gauss's theorem is then used to convert the volume integral into a surface integral over the boundary $\partial{\cal{V}}$ with area element $\left|d^{3}A\right|$.
\begin{eqnarray} &0=\int_{\partial{\cal{V}}}\rho_{0}u^{\mu}n_{\mu}\left|d^{3}A\right| \\
& 0= \left[\int^{r+\Delta r}_{r}\int^{2\pi}_{0}\int^{+h}_{-h}\textrm{d}r\textrm{d}\phi\textrm{d}z\; \sqrt{-{\bf{g}}} \rho_{0}u^{t}  \right]^{t=t_{0}+T}_{t=t_{0}} \nonumber\\
&\quad\;+\left[ \int^{t+T}_{t}\int^{2\pi}_{0}\int^{+h}_{-h}\textrm{d}t\textrm{d}\phi\textrm{d}z\;  \sqrt{-{\bf{g}}}\rho_{0}u^{r}  \right]^{r'=r+\Delta r}_{r'=r} \nonumber \\
&\quad\; +\left[\int^{t_{0}+T}_{t}\int^{r+\Delta r}_{r}\int^{+h}_{-h}\textrm{d}t\textrm{d}r\textrm{d}z\; \sqrt{-{\bf{g}}}\rho_{0}u^{\phi}  \right]^{\phi=2\pi}_{\phi=0}\nonumber \\
&\quad\; +\left[\int^{t_{0}+T}_{t_{0}}\int^{r+\Delta r}_{r}\int^{2\pi}_{0}\textrm{d}t\textrm{d}r\textrm{d}\phi\; \sqrt{-{\bf{g}}}\rho_{0}u^{z} \right]^{z=+h}_{z=-h} \label{eq:massconservation}
\end{eqnarray}

The first and third terms in the above expression are zero by the assumed stationarity and axisymmetry of the system. The final term is also zero because there is no motion in the vertical direction, $u^{z}=0$. Therefore Eq.\ \ref{eq:massconservation} simplifies to
\begin{eqnarray}
&0=2\pi T\Delta r \left( \sqrt{-{\bf{g}}}\Sigma(r) u^{r}\right)_{,r} \nonumber\\
& \Rightarrow \dot{M}_{0}=-2\pi\sqrt{-{\bf{g}}} \Sigma(r) u^{r}=\textrm{constant} \, ,
\end{eqnarray}
where $\dot{M}_{0}$ is the accretion rate. 

The second conservation law is that of angular momentum. Again the differential form of the conservation law is integrated over the volume ${\cal{V}}$ and Gauss's law used to turn this into a surface integral over $\partial{\cal{V}}$.
\begin{eqnarray}
&0=&\nabla_{\mu}J^{\mu}\quad \textrm{where} \quad J^{\mu}=T^{\mu\nu}\left(\frac{\partial}{\partial\phi}\right)_{\nu}\nonumber\\
&0=&\int_{\partial{\cal{V}}}J^{\mu}n_{\mu} \left|d^{3}A\right|\nonumber \\
\end{eqnarray}

\begin{widetext}
\begin{eqnarray}\label{eq:angmomconserv}
&0=&\left[\int^{t_{0}+T}_{t_{0}}\int^{2\pi}_{0}\int^{+h}_{-h}\textrm{d}t\textrm{d}\phi\textrm{d}z\; \left[\rho_{0}(1+\Pi)u_{\phi}u^{r} + t_{\phi}^{\;r} +u_{\phi}q^{r} + q_{\phi}u^{r} \right] \sqrt{-{\bf{g}}} \right]^{r'=r+\Delta r}_{r'=r}\nonumber\\
&&\quad\quad +\left[\int_{t_{0}}^{t_{0}+\Delta t}\int_{r}^{r+\Delta r}\int_{0}^{2\pi}\textrm{d}t\textrm{d}r\textrm{d}\phi\; \left[ \rho_{0}(1+\Pi)u_{\phi}u^{z} + t_{\phi}^{\;z}+u_{\phi}q^{z}+q_{\phi}u^{z} \right] \sqrt{-{\bf{g}}}  \right]^{z=+h}_{z=-h} 
\end{eqnarray}\end{widetext}
The $\phi$ index has been lowered and the $t$ and $\phi$ integral terms have been set to zero due to the assumed stationarity and axisymmetry of the system. Using the negligible internal energy condition derived above, Eqs.\ \ref{eq:heatflow} and \ref{eq:tcomp}, and the fact that $u^{z}=0$ this becomes
\begin{eqnarray}  
&0=&\left[2\pi T\int_{-h}^{+h}\textrm{d}z\; \left[\rho_{0}u_{\phi}u^{r} + t_{\phi}^{\;r}\right] \sqrt{-{\bf{g}}} \right]^{r'=r+\Delta r}_{r'=r}\nonumber\\
&&+\left[2\pi T \int_{r}^{r+\Delta r}\textrm{d}r\; u_{\phi}q^{z} \sqrt{-{\bf{g}}}  \right]^{z=+h}_{z=-h} \\
&\Rightarrow&4\pi\sqrt{-{\bf{g}}}F(r)\tilde{L}=\left[\dot{M}_{0}\tilde{L}-2\pi \sqrt{-{\bf{g}}}W_{\phi}^{\; r}\right]_{,r} \; .\label{eq:diff1}
\end{eqnarray}
An extra factor of two has appeared on the left-hand side of Eq.\ \ref{eq:diff1} from the fact that a flux $F(r)$ is radiated from both sides of the disk.

The third and final conservation law is that of conservation of energy. By performing the same type of manipulations to this equation as was done for Eq.\ \ref{eq:angmomconserv} we obtain,
\begin{eqnarray} & 0=\nabla_{\mu}E^{\mu}\quad\textrm{where}\quad E^{\mu}=-T^{\mu\nu}\left(\frac{\partial}{\partial t}\right)_{\nu}\; , \\
&\left[ \dot{M}_{0}\tilde{E}+2\pi \sqrt{-{\bf{g}}}W_{t}^{\; r} \right]_{,r} = 4\pi \sqrt{-{\bf{g}}}F(r)\tilde{E} \; .\end{eqnarray}
Making use of the orthogonality $u^{\mu}t_{\mu}^{\; \nu}=0$ which implies that $u^{\mu}W_{\mu}^{\;\nu}=0\;\Rightarrow\; W_{t}^{\;r}+\Omega W_{\phi}^{\;r}=0$, this equation may be rewritten in terms of $W_{\phi}^{\; r}$ as was the case with the angular momentum equation.
\begin{equation}\label{eq:diff2} \left[ \dot{M}_{0}\tilde{E}-2\pi \sqrt{-{\bf{g}}}W_{\phi}^{\; r}\Omega \right]_{,r} = 4\pi \sqrt{-{\bf{g}}}F(r)\tilde{E} \end{equation}

From Eqs.\ \ref{eq:En}, \ref{eq:Lz} and \ref{eq:omega} it can be seen that the energy, angular momentum and angular velocity satisfy the following energy angular momentum relation,
\begin{equation}\label{eq:ELOmega} \tilde{E}_{,r}=\Omega \tilde{L}_{,r} \; .\end{equation}

Eqs.\ \ref{eq:diff1} and \ref{eq:diff2} may now be integrated to find the radial dependence of the flux. This is done by multiplying Eq.\ \ref{eq:diff1} by $\Omega$ and subtracting the result from Eq.\ \ref{eq:diff2} to obtain an expression for the torque;
\begin{equation} W_{\phi}^{\; r} = 2 F(r) \frac{\Omega \tilde{L}-\tilde{E}}{\Omega _{,r}}\, . \end{equation}
Substituting this back into Eq.\ \ref{eq:diff1} and rearranging and using the energy angular momentum relation in Eq.\ \ref{eq:ELOmega} gives a differential equation for $F(r)$,
\begin{equation} \left[ 4\pi \sqrt{-{\bf{g}}}\frac{\left(\tilde{E}-\Omega\tilde{L}\right)^{2}}{\Omega_{,r}} F(r) \right]_{,r} = \dot{M}_{0}\left(\tilde{E}-\Omega\tilde{L}\right)\tilde{L}_{,r}\, , \end{equation}
which may be readily integrated. To fix the constant of integration we use the zero torque boundary condition at the inner edge of the disk, $F(r_{\textrm{isco}})=0$. Therefore we have an expression for the radial flux from the disk,
\begin{equation}\label{eq:radialflux} F(r)=\frac{-\dot{M}_{0}\Omega_{,r}}{4\pi \sqrt{-{\bf{g}}}\left( \tilde{E}-\Omega \tilde{L} \right)^{2}} \int_{r_{\textrm{isco}}}^{r}\left(\tilde{E}-\Omega \tilde{L}\right) L_{,r} \textrm{d}r \, .\end{equation}

For completeness we also present the final radial structure expression derived in \cite{1974ApJ...191..499P}, the expression for the torque per unit circumference as a function of radius, $W_{\phi}^{r}$,
\begin{eqnarray} &W_{\phi}^{r}(r)&=\frac{-\dot{M}_{0}\Omega_{,r}}{2\pi \sqrt{-{\bf{g}}}\left( \tilde{E}-\Omega \tilde{L} \right)^{2}} \frac{\tilde{E}-\Omega \tilde{L}}{-\Omega_{,r}}\nonumber \\
&&\quad\times\int_{r_{\textrm{isco}}}^{r}\left(\tilde{E}-\Omega \tilde{L}\right) L_{,r} \textrm{d}r  \, .\end{eqnarray}

\clearpage

\bibliography{bibliography}

\end{document}